\documentclass[published]{JHEP3} 
\JHEP{00(2006)000}
\JHEPspecialurl{http://jhep.sissa.it/JOURNAL/JHEP3.tar.gz}

\usepackage{epsfig}

\def\bea{\begin{eqnarray}}
\def\eea{\end{eqnarray}}
\def\bec{\begin{center}}
\def\ec{\end{center}}

\def\beq{\begin{equation}}
\def\eeq{\end{equation}}

\title{Neutralino Dark Matter in Mirage Mediation}

\author{Kiwoon Choi,~Kang Young Lee,~Yasuhiro Shimizu\\
Department of Physics, KAIST, Daejeon 305-701, Korea \\
        E-mail: \email{kchoi@muon.kaist.ac.kr},
         \email{kylee@muon.kaist.ac.kr},
         \email{shimizu@muon.kaist.ac.kr} }

\author{Yeong Gyun Kim \\
Astrophysical Research Center for
the Structure and Evolution of the Cosmos,\\
Sejong University,
Seoul 143-747, Korea \\
        E-mail: \email{ygkim@muon.kaist.ac.kr} }

\author{Ken-ichi Okumura\\
Department of Physics, Kyushu University, Fukuoka 812-8581, Japan \\
        E-mail: \email{okumura@higgs.phys.kyushu-u.ac.jp} }

\received{} 
\revised{}
\accepted{}           

\preprint{KAIST-TH 2006/08}
\preprint{KYUSHU-HET-96}

\abstract{We study the phenomenology of neutralino dark matter
(DM) in mirage mediation scenario  of supersymmetry breaking which
results from the moduli stabilization in some string/brane models.
Depending upon the model parameters, especially  the anomaly to
modulus mediation ratio determined by the moduli stabilization
mechanism, the nature of the lightest supersymmetric particle
(LSP) changes from Bino-like neutralino to Higgsino-like one via
Bino-Higgsino mixing region. For the Bino-like LSP, the standard
thermal production mechanism can give a  right amount of relic DM
density through the stop/stau-neutralino coannihilation or the
pseudo-scalar Higgs resonance process. We also examine the
prospect of direct and indirect DM detection in various parameter
regions of mirage mediation. Neutralino DM in galactic halo might
be detected by near future direct detection experiments in the
case of Bino-Higgsino mixed LSP. The gamma ray flux from Galactic
Center might  be detectable also if  the DM density profile takes
a cuspy shape.}


\dedicated{}

\begin{document}

\section{Introduction}

Low energy supersymmetry (SUSY) is a promising candidate for
physics beyond the standard model at TeV scale
\cite{Nilles:1983ge}. In addition to solving the naturalness
problem associated with the weak to Planck scale hierarchy $M_{\rm
weak}/M_{Pl}\sim 10^{-16}$,
low energy SUSY  provide a natural candidate for the cold dark
matter (DM) in the universe \cite{lspdm}, the neutralino being the
lightest supersymmetric particle (LSP), under the assumption of
$R$-parity conservation. The assumption that the LSP neutralino
constitutes DM constrains SUSY model in various ways. The first
constraint comes from the requirement that the LSP of the model
should be a neutralino. Although not severe, the current limits on
the direct or indirect detection of DM also provide an additional
constraint on the model. Finally, if the mechanism of cosmological
neutralino production is specified, the model parameters should be
in the range giving the correct DM abundance measured by WMAP
\cite{wmap}: \bea 0.085< \Omega_{DM}h^2< 0.119. \eea

Recent development in  string moduli stabilization \cite{gkp,kklt}
has led to a new pattern of soft SUSY breaking terms which has not
been explored before \cite{choi1}. In KKLT-type moduli
stabilization scenario \cite{kklt}, the volume modulus $T$ is
stabilized at SUSY AdS vacuum by non-perturbative effect,
and this SUSY  AdS vacuum of $T$ is lifted to a phenomenologically
viable dS or Minkowski vacuum by SUSY breaking brane.
  Such set-up
leads to a mass pattern \cite{choi1}: \bea \frac{F^T}{T} \,\sim\,
\frac{m_{3/2}^2}{m_T}\,\sim\,
\frac{m_{3/2}}{\ln(M_{Pl}/m_{3/2})},\eea where $m_T$ and $F^T$
denote the modulus mass and $F$-component, respectively. If the
SUSY breaking brane is sequestered from the visible sector, the
soft masses of visible fields are determined by  the modulus
mediation \cite{modulus} of ${\cal O}(F^T/T)$ and the anomaly
mediation \cite{anomaly} of ${\cal O}(m_{3/2}/8\pi^2)$ which are
comparable to each other if the gravitino mass $m_{3/2}\sim 10$
TeV as required to give the weak scale size of soft masses. An
interesting consequence of this mixed mediation is that soft
masses are unified at a mirage messenger scale \cite{Choi:2005uz}
\bea M_{\rm
mir}=M_{GUT}\left(\frac{m_{3/2}}{M_{Pl}}\right)^{\alpha/2},\eea
 where
$\alpha$ is a parameter of order unity  which represents the
anomaly to modulus mediation ratio\footnote{Here we are following
the notations of Ref.~\cite{Choi:2005uz}. We warn the readers that
some subsequent works \cite{falkowski05,baer06} are using a
differently defined $\alpha$ which corresponds to $\alpha_{\rm
Ref.\cite{falkowski05}}=16\pi^2/[\alpha_{\rm
ours}\ln(M_{Pl}/m_{3/2})]\simeq 4.9/\alpha_{\rm ours}$. We stress
that our definition of $\alpha$ is more convenient for  matching
the mirage mediation to underlying moduli stabilization model. For
instance, the original KKLT moduli stabilization gives
$\alpha_{\rm ours}=1$, and many of its generalizations give a
rational value of $\alpha_{\rm ours}$.}. This feature named as
mirage mediation \cite{ratz} is due to the particular property of
anomaly mediation which is closely related to the renormalization
group (RG) evolution of soft parameters. KKLT-type moduli
stabilization typically gives a positive $\alpha={\cal O}(1)$, and
thus a mirage messenger scale hierarchically lower than $M_{GUT}$.
The corresponding low energy superparticle masses have a quite
different pattern from those in other SUSY breaking scenarios such
as mSUGRA, gauge mediation and anomaly mediation
\cite{Choi:2005uz,endo05,falkowski05,baer06}.

In this paper, we wish to study  some  phenomenological aspects of
neutralino DM in mirage mediation, particularly examine the
parameter range giving LSP neutralino, the relic DM density under
the assumption of standard thermal production, and the prospect of
direct or indirect detection of DM neutralino. As we will see,
when the anomaly to modulus mediation ratio $\alpha$ increases
from zero to a positive value of order unity, or equivalently the
mirage messenger scale $M_{\rm
mir}=M_{GUT}(m_{3/2}/M_{Pl})^{\alpha/2}$ decreases from $M_{GUT}$
to a hierarchically lower scale, the nature of LSP neutralino is
changed from Bino-like to Higgsino-like via Bino-Higgsino mixing
region. The enhanced Higgsino component can be understood by the
gluino, Wino and Bino mass ratios in mirage mediation: \bea
M_3:M_2:M_1\simeq
(1-0.3\alpha)g_3^2:(1+0.1\alpha)g_2^2:(1+0.66\alpha)g_1^2,\eea
where $\alpha\neq 0$ represents  the effects of anomaly mediation.
Thus, compared to the mSUGRA-type pure modulus mediation
($\alpha=0$), the gluino to Bino mass ratio for $\alpha\sim 1$ is
significantly reduced. This results in smaller $|m_{H_u}^2|/M^2_1$
 at the weak scale,
 and thus a smaller Higgsino to Bino mass
ratio when the electroweak symmetry breaking condition is imposed.
(Here $m^2_{H_u}$ is the soft SUSY breaking mass-square of the
up-type Higgs doublet $H_u$.) As a consequence, for $\alpha\sim
1$, the LSP neutralino has a sizable Higgsino component over a
large fraction of the parameter space, and this Higgsino component
eventually becomes dominant for a larger value of $\alpha$.  For
the parameter region leading to Bino-like LSP, the thermal
production mechanism can give a right amount of relic DM density
through the stop/stau-neutralino coannihilation process or the
pseudo-scalar Higgs resonance effect, depending upon the value of
$\tan\beta=\langle H_u\rangle/\langle H_d\rangle$. In overall,
compared to mSUGRA scenario, a significantly larger fraction of
the parameter space of mirage mediation can give the WMAP DM
density under the assumption of the conventional thermal
production of neutralino LSP, while satisfying all known
phenomenological constraints.  As for the DM detection, neutralino
DM in galactic halo might be detected by near future direct
detection experiments in the case of Bino-Higgsino mixed LSP which
can be  obtained over a large fraction of the parameter space. The
gamma ray flux from galactic center might be detectable also if
the DM density profile takes a cuspy shape.

 Some phenomenological and
cosmological aspects of mirage mediation have been investigated by
several groups
\cite{Choi:2005uz,endo05,falkowski05,baer06,baer,yama}. In
particular, the properties of neutralino DM have been studied in
Ref.~\cite{baer06}.
The sub-GUT CMSSM model of Ref.~\cite{ellis06} also shares a
common feature with  mirage mediation as it assumes the
unification scale of soft parameters below the GUT scale. In this
paper, we provide a more extensive analysis of  neutralino DM for
certain range of model parameters which are expected to be
obtained in KKLT-type moduli stabilization scenarios together with
a more detailed study of  the direct and indirect DM detections.
 Our results agree qualitatively
well with Ref.~\cite{baer06} when the considered parameter space
overlaps.

Recent study of moduli cosmology \cite{yama} in KKLT-type moduli
stabilization
 indicates that if the early universe underwent a period
dominated by the coherent oscillation of moduli, there can be too
many gravitinos and neutralino LSPs produced by moduli decays,
which would spoil the Big-Bang nucleosynthesis or overclose the
universe. This requires a mechanism to dilute the primordial
moduli oscillation, e.g. a thermal inflation as proposed in
\cite{stewart}. In the presence of such thermal inflation, the
neutralino DM can be produced either by the conventional  thermal
production mechanism or by the decays of flaton triggering thermal
inflation, depending upon the details of reheat procedure.
Throughout this paper, we will focus on the relic density of
neutralino DM produced  by the conventional thermal production
mechanism while leaving an additional non-thermal production
\cite{yama,nonthermal} as an open possibility.

In fact, if the anomaly to modulus mediation ratio $\alpha$ takes
a value giving a mirage messenger scale $M_{\rm
mir}=M_{GUT}(m_{3/2}/M_{Pl})^{\alpha/2}$  much lower than the
intermediate scale, the squark/slepton mass-squares have negative
values at certain {\it high} renormalization point $\mu\gtrsim
\mu_c$, although they become positive at lower renormalization
point around TeV \cite{Choi:2005uz,falkowski05}. As long as the
low energy squark/slepton mass-squares are positive\footnote{It
has been argued that a negative stop mass-square at high
renormalization point is helpful for ameliorating the fine-tuning
for the electroweak symmetry breaking in the MSSM \cite{hdkim}.},
the model has a correct color/charge preserving (but electroweak
symmetry breaking) vacuum with squark/slepton fields $\phi=0$.
Still, tachyonic squark/slepton mass-squares at high
renormalization point $\mu\gtrsim \mu_c$ might give a deeper
color/charge breaking (CCB) minimum or a unbounded from below
(UFB) direction in the effective potential at $\phi
> \mu_c$ \cite{munoz}, which would make the color/charge conserving vacuum metastable.
 In the presence of such CCB minimum or UFB direction, there are two
points to be clarified to make sure that the model is
phenomenologically viable. One first needs a cosmological scenario
which allows our universe to be settled down at the correct
vacuum, not at the CCB minimum or UFB direction. The second is
that the color/charge preserving vacuum should be stable enough
against the tunnelling into CCB minimum or UFB direction. The
first point might depend on the detailed history of the early
universe. However in view of that squark/sleptons get large
positive mass-squares in the high temperature limit, it is a
rather plausible assumption that squark/sleptons are settled down
at the color/charge preserving minimum after the inflation
\cite{kuzenko}. As for the vacuum stability, it has been noticed
that the potential barrier between $\phi=0$ and $\phi\gtrsim
\mu_c$ gives a tunnelling rate much less than the Hubble expansion
rate as long as $\mu_c\gtrsim 10$ TeV \cite{riotto}. In mirage
mediation with positive $\alpha={\cal O}(1)$, once one requires a
successful electroweak symmetry breaking, $\mu_c$ is  higher than
$10^3$ TeV, thus satisfies safely the stability condition.  In
this paper, we do not take the existence of CCB minimum or UFB
direction at large squark/slepton value $|\phi|\gg 10$ TeV as a
serious problem of the model as long as a good color/charge
preserving and electroweak symmetry breaking vacuum exists, and
focus on the phenomenology of the model under the assumption that
we are living in the color/charge preserving local minimum which
is stable enough to have a lifetime longer than the age of the
universe.


The organization of this paper is as follows. In section II, we
review the basic features of mirage mediation. In section III, we
examine the prospect of neutralino DM in  intermediate scale
mirage mediation scenario  for several different choices of the
matter/Higgs modular weights. In section IV, we extend the
analysis to general values of the mirage messenger scale. Section
V is the conclusion, and Appendix A contains a summary of our
convention and notation.

\section{Mirage mediation}

Mirage mediation is a natural consequence of the KKLT-type moduli
stabilization scenario satisfying the following two  assumptions:
(i) the modulus $T$ (or dilaton) which determines the standard
model gauge couplings is stabilized by non-perturbative effects
and (ii) SUSY is broken by a brane-localized source which is
sequestered from the visible sector. A well known example of such
set-up is the KKLT moduli stabilization \cite{kklt} in type IIB
string theory\footnote{As was noticed in \cite{choi3,hebecker},
due to the effect of throat vector multiplet, the sequestering
might not be precise enough in the case of KKLT compactification
of type IIB string theory. The size of non-sequestered soft scalar
mass induced by the exchange of throat vector multiplet is quite
sensitive to the unknown details of compactification, however
there exists a reasonable parameter limit in which the
non-sequestered effects can be safely ignored \cite{choi3}.} A
similar but simpler example would be 5D brane model with a flat
interval in one side and an warped interval in other side, in
which  SUSY breaking brane is introduced at the IR fixed point of
the warped interval.

Under these two assumptions, the 4D effective action of the
visible sector fields and the gauge coupling modulus $T$ is given
by
 \bea
 \label{superspace}
 \int d^4\theta \left[-3CC^*e^{-K/3}
-C^2C^{*2}{\cal P}_{\rm
lift}\theta^2\bar{\theta}^2\right]+\left(\int d^2\theta \left[
\frac{1}{4}f_aW^{a\alpha}W^a_\alpha+C^3W\right]+{\rm h.c.}
\right), \eea where $C=C_0+F^C\theta^2$ is the chiral compensator
superfield, $f_a$ are the holomorphic gauge kinetic function of
the visible sector gauge fields, $K$ and $W$ are the effective
K\"ahler potential and superpotential of the visible matter
superfields $\Phi_i$ and the gauge coupling modulus $T$, which
would be obtained by integrating out heavy moduli. As long as the
SUSY breaking brane is sequestered from the visible gauge and
matter superfields, its low energy consequence can be described by
a simple spurion operator of the form ${\cal P}_{\rm
lift}\theta^2\bar{\theta}^2$, independently of the detailed
 dynamics on the SUSY-breaking brane. Assuming an axionic shift
symmetry: \bea \label{shift} U(1)_T:\quad {\rm Im}(T)\rightarrow
{\rm Im}(T)+\mbox{real constant}\eea which is broken by
non-perturbative term in the superpotential, the model is given by
\bea
\label{u1tmodel}K&=&K_0(T+T^*)+Z_i(T+T^*)\Phi^*_i\Phi_i, \nonumber \\
W &=&  w-Ae^{-aT}+\frac{1}{6}\lambda_{ijk} \Phi_i\Phi_j\Phi_k,
\nonumber
\\
f_a&=& kT+\Delta f_a, \nonumber \\
{\cal P}_{\rm lift}&=& {\cal P}_{\rm lift}(T+T^*), \eea where $a$
and $k$ are (discrete) real constants, while $\Delta f_a, w, A$
and $\lambda_{ijk}$ are  complex effective constants obtained
after heavy moduli are integrated out. The axionic symmetry
(\ref{shift}) ensures that $K_0$,
 $Z_i$ and ${\cal P}_{\rm lift}$ depend only on $T+T^*$, and $a$ and $k$  are real constant.
With these features, the resulting gaugino masses and trilinear
$A$ parameters preserve CP as was pointed out in \cite{susycp}.

There might be various ways to generate the modulus superpotential
$W_0=w-Ae^{-aT}$ stabilizing  $T$. Generically the
non-perturbative term $e^{-aT}$ can be induced by either  a
gaugino condensation of $T$-dependent hidden gauge interaction or
a stringy instanton whose Euclidean action is controlled by $T$.
As for the constant term $w$, it might be induced by a fine-tuned
configuration of fluxes as in the original KKLT scenario
\cite{kklt}, or alternatively by $T$-independent non-perturbative
effect whose strength is controlled by heavy moduli \cite{luty}.
As we will see, the non-perturbative stabilization of $T$ by $W_0$
generates a little hierarchy between the modulus mass and the
gravitino mass: \bea \frac{m_T}{m_{3/2}}\,\sim\, aT\,\sim\,
\ln(M_{Pl}/m_{3/2}), \eea which results in a little suppression of
the modulus $F$-component: \bea \frac{F^T}{T} \,\sim
\,\frac{m_{3/2}^2}{m_T} \,\sim\,
\frac{m_{3/2}}{\ln(M_{Pl}/m_{3/2})}.\eea Then,  $F^T/T$ naturally
has a size comparable to the anomaly mediated soft mass of ${\cal
O}(m_{3/2}/4\pi^2)$ for the gravitino mass around TeV. If the SUSY
breaking source is sequestered from the visible sector, the soft
terms of visible fields are determined by the modulus mediation of
${\cal O}(F^T/T)$ and the anomaly mediation of ${\cal
O}(m_{3/2}/4\pi^2)$ which are comparable to each other. This leads
to a mirage unification \cite{Choi:2005uz} of soft masses at a
scale hierarchically different from the gauge coupling unification
scale $M_{GUT}$.

In the Einstein frame, the modulus potential from
(\ref{superspace}) is given by
 \bea
 \label{moduluspotential}
 V_{\rm TOT}
=e^{K_0}\left[(\partial_T\partial_{\bar{T}}K_0)^{-1}|D_TW_{0}|^2-3|W_{0}|^2\right]+V_{\rm
lift}, \eea where
\bea
 W_0=w-Ae^{-aT}, \quad V_{\rm lift}= e^{2K_0/3}{\cal P}_{\rm
lift}.\eea The superspace lagrangian density (\ref{superspace})
also determines the auxiliary components of $C$ and $T$ as
\bea
\label{approx-F} \frac{F^C}{C_0}&=
&\frac{1}{3}\partial_TK_0F^T+m_{3/2}^*,
\nonumber \\
F^T&=
&-e^{K_0/2}\left(\partial_T\partial_{T^*}K_0\right)^{-1}\left(D_T
W_{0}\right)^*, \eea where $m_{3/2}= e^{K_0/2}W_0$. Note that one
can always make both $w$ and $A$ real by appropriate $U(1)_R$ and
$U(1)_T$ transformations. In such field basis, the $U(1)_T$
invariance of $K_0$ assures that both $m_{3/2}$ and $F^T$ are
real. In the following, we will use this field basis in which the
CP invariance of soft parameters is easier to be recognized.

To stabilize $T$ at a reasonably large value while having
$m_{3/2}$ hierarchically smaller than $M_{Pl}$, one needs to
assume that $w$  is hierarchically smaller than $A$ in the unit
with $M_{Pl}=1$. Since $w\sim m_{3/2}$ and one needs $m_{3/2}\sim
10$ TeV to get the weak scale superparticle masses, $\ln(A/w)$
typically has a value of ${\cal O}(4\pi^2)$. It is then
straightforward to compute the vacuum values of $T$ and $F^T$ by
minimizing the modulus potential (\ref{moduluspotential}) under
the fine tuning condition $\langle V_{\rm TOT}\rangle =0$. At
leading order in $\epsilon=1/\ln(A/w)$,
one finds \cite{choi1} \bea \label{vev} a T &\simeq &\ln(A/w),\nonumber \\
 \frac{F^T}{T+T^*}
 &\simeq& \frac{m_{3/2}}{\ln(A/w)}\left(1+\frac{3\partial_T\ln({\cal P}_{\rm
lift})}{2\partial_TK_0}
 \right)\,,\eea
which shows that $F^T/T$ is indeed of the order of
$m_{3/2}/4\pi^2$ for $\ln(A/w)\sim 4\pi^2$.

Let us consider the soft terms of canonically normalized visible
fields derived from the 4D effective action (\ref{superspace}):
\begin{eqnarray}
{\cal L}_{\rm
soft}&=&-\frac{1}{2}M_a\lambda^a\lambda^a-\frac{1}{2}m_i^2|\phi_i|^2
-\frac{1}{6}A_{ijk}y_{ijk}\phi_i\phi_j\phi_k+{\rm h.c.},
\end{eqnarray}
where $\lambda^a$ are gauginos, $\phi_i$ are the scalar component
of $\Phi_i$ and $y_{ijk}$ are the canonically normalized Yukawa
couplings: \bea
y_{ijk}=\frac{\lambda_{ijk}}{\sqrt{e^{-K_0}Z_iZ_jZ_k}}. \eea For
$F^T/T\sim m_{3/2}/4\pi^2$,
 the soft parameters  at energy scale just below  $M_{GUT}$ are
 determined by the modulus-mediated and
 anomaly-mediated contributions which are comparable to each
 other.
 One then finds \cite{choi1}
\begin{eqnarray}
\label{soft1} M_a&=& M_0 +\frac{m_{3/2}}{16\pi^2}\,b_ag_a^2,
\nonumber \\
A_{ijk}&=&\tilde{A}_{ijk}-
\frac{m_{3/2}}{16\pi^2}\,(\gamma_i+\gamma_j+\gamma_k),
\nonumber \\
m_i^2&=& \tilde{m}_i^2-\frac{m_{3/2}}{16\pi^2}M_0\,\theta_i
-\left(\frac{m_{3/2}}{16\pi^2}\right)^2\dot{\gamma}_i
\label{eq:bc}
\end{eqnarray}
where $M_0$, $\tilde{A}_{ijk}$ and $\tilde{m}_i$ are the pure
modulus-mediated gaugino mass, trilinear $A$-parameters and
sfermion masses which are given by
\bea
\label{tmediation} M_0&=&F^T\partial_T\ln{\rm Re}(f_a),
\nonumber \\
\tilde{m}_i^2 &=& -F^TF^{T*}\partial_T\partial_{\bar{T}}
\ln(e^{-K_0/3}Z_i),
\nonumber \\
\tilde{A}_{ijk}&=& -F^T\partial_T\ln
             \left(\frac{\lambda_{ijk}}{e^{-K_0}Z_iZ_jZ_k}\right)
             \,=\,
             F^T\partial_T\ln(e^{-K_0}Z_iZ_jZ_k)
\nonumber \\
              &=&\tilde{A}_i+\tilde{A}_j+\tilde{A}_k
               \quad\,
               \mbox{for}\quad\,\tilde{A}_i=F^T\partial_T\ln(e^{-K_0/3}Z_i).
\eea Here we have used that the holomorphic Yukawa couplings
$\lambda_{ijk}$ are $T$-independent constants as a consequence of
the axionic  shift symmetry $U(1)_T$, and the one-loop beta
function coefficient  $b_a$, the anomalous dimension $\gamma_i$
and its derivative $\dot{\gamma}_i$, and $\theta_i$ are defined as
\bea
b_a&=&-3{\rm tr}\left(T_a^2({\rm Adj})\right)
        +\sum_i {\rm tr}\left(T^2_a(\phi_i)\right),
\nonumber \\
\gamma_i&=&2\sum_a g^2_a C^a_2(\phi_i)-\frac{1}{2}\sum_{jk}|y_{ijk}|^2,
\nonumber \\
\dot{\gamma}_i&=&8\pi^2\frac{d\gamma_i}{d\ln\mu},\nonumber \\
\theta_i&=& 4\sum_a g^2_a C^a_2(\phi_i)-\sum_{jk}|y_{ijk}|^2
\frac{\tilde{A}_{ijk}}{M_0},
 \eea where the
quadratic Casimir $C^a_2(\phi_i)=(N^2-1)/2N$ for a fundamental
representation $\phi_i$ of the gauge group $SU(N)$,
$C_2^a(\phi_i)=q_i^2$ for the $U(1)$ charge $q_i$ of $\phi_i$, and
$\omega_{ij}=\sum_{kl}y_{ikl}y^*_{jkl}$ is assumed to be diagonal.
In Appendix A, we provide a summary of our convention and
notation.

For our later discussion, it is convenient to define
\bea
\alpha\,\equiv\,\frac{m_{3/2}}{M_0\ln(M_{Pl}/m_{3/2})},\quad
 a_i\,\equiv\,\frac{\tilde{A}_{i}}{M_0}, \quad
 c_i\,\equiv\, \frac{\tilde{m}_i^2}{M_0^2},
\label{eq:def}
 \eea
 where $\alpha$ represents the anomaly to modulus mediation
 ratio, while $a_{i}$ and $c_i$ parameterize the pattern of the pure modulus mediated soft masses.
Then the boundary values of soft parameters at $M_{GUT}$ are given
by
\begin{eqnarray}
M_a&=& M_0 \Big[\,1+\frac{\ln(M_{Pl}/m_{3/2})}{16\pi^2} b_a
g_a^2\alpha\,\Big],\nonumber \\
A_{ijk}&=&M_0\Big[\,(a_i+a_j+a_k)
-\frac{\ln(M_{Pl}/m_{3/2})}{16\pi^2}(\gamma_i+\gamma_j+\gamma_k)\alpha\,\Big],
\nonumber \\
m_i^2&=&M_0^2\Big[\,c_i-\,\frac{\ln(M_{Pl}/m_{3/2})}{16\pi^2}
\theta_i\alpha-\left(\frac{\ln(M_{Pl}/m_{3/2})}{16\pi^2}\right)^2\dot{\gamma}_i\alpha^2\,\Big],
\label{eq:bc1}
\end{eqnarray}
where \bea
\theta_i=4\sum_a g^2_a C^a_2(\phi_i)-\sum_{jk}|y_{ijk}|^2(a_i+a_j+a_k).\eea
In this prescription, generic mirage mediation is parameterized by
\bea M_0,\,\, \alpha,\,\, a_i,\,\, c_i,\,\, \tan\beta,\eea where
we have replaced the Higgs mass parameters  $\mu$ and $B$ by
$\tan\beta=\langle H_u\rangle/\langle H_d\rangle$ and $M_Z$ as
usual. As we will see, this parameterization of mirage mediation is
particularly convenient when one compute the mirage mediation
parameters from underlying SUGRA model. For instance, $\alpha$,
$a_i$ and $c_i$ are given by simple rational numbers in minimal
KKLT-type moduli stabilization.

Taking into account the 1-loop RG evolution, the soft masses of
(\ref{soft1}) at $M_{GUT}$ leads to low energy soft masses which
can be described in terms of the mirage messenger scale: \bea
M_{\rm mir}=\frac{M_{GUT}}{(M_{Pl}/m_{3/2})^{\alpha/2}}. \eea For
instance, the low energy gaugino masses are given by
\cite{Choi:2005uz}\bea \label{lowgaugino} M_a(\mu)=M_0\left[\,
1-\frac{1}{8\pi^2}b_ag_a^2(\mu)\ln\left(\frac{M_{\rm
mir}}{\mu}\right)\,\right] =\frac{g_a^2(\mu)}{g_a^2(M_{\rm
mir})}M_0, \eea showing that the gaugino masses are unified at
$M_{\rm mir}$, while the gauge couplings are unified at $M_{GUT}$.
The low energy values of $A_{ijk}$ and $m_i^2$ generically depend
on the associated Yukawa couplings $y_{ijk}$. However if $y_{ijk}$
are small enough or \bea\label{con1} a_i+a_j+a_k=c_i+c_j+c_k=1
\quad \mbox{for} \quad y_{ijk}\sim 1,\eea their low energy values
are given by \cite{Choi:2005uz}\bea \label{lowsfermion}
A_{ijk}(\mu)&=& M_0\left[\,a_i+a_j+a_k+
\frac{1}{8\pi^2}(\gamma_i(\mu)
+\gamma_j(\mu)+\gamma_k(\mu))\ln\left(\frac{M_{\rm
mir}}{\mu}\right)\,\right], \nonumber \\
m_i^2(\mu)&=&M_0^2\left[\,c_i-\frac{1}{8\pi^2}Y_i\left(
\sum_jc_jY_j\right)g^2_Y(\mu)\ln\left(\frac{M_{GUT}}{\mu}\right)\right.
\nonumber \\
&+&\left.\frac{1}{4\pi^2}\left\{
\gamma_i(\mu)-\frac{1}{2}\frac{d\gamma_i(\mu)}{d\ln\mu}\ln\left(
\frac{M_{\rm mir}}{\mu}\right)\right\}\ln\left( \frac{M_{\rm
mir}}{\mu}\right)\,\right], \eea where $Y_i$ is the $U(1)_Y$
charge of $\phi_i$. Quite often, the modulus-mediated squark and
slepton masses have a common value, i.e.
$c_{\tilde{q}}=c_{\tilde{\ell}}$. Then, according to the above
expression of low energy sfermion mass, the 1st and 2nd generation
squark and slepton masses are unified again at the mirage
messenger scale $M_{\rm mir}$.

\begin{figure}[ht!]
\begin{center}
\includegraphics[height=7cm,width=7cm]{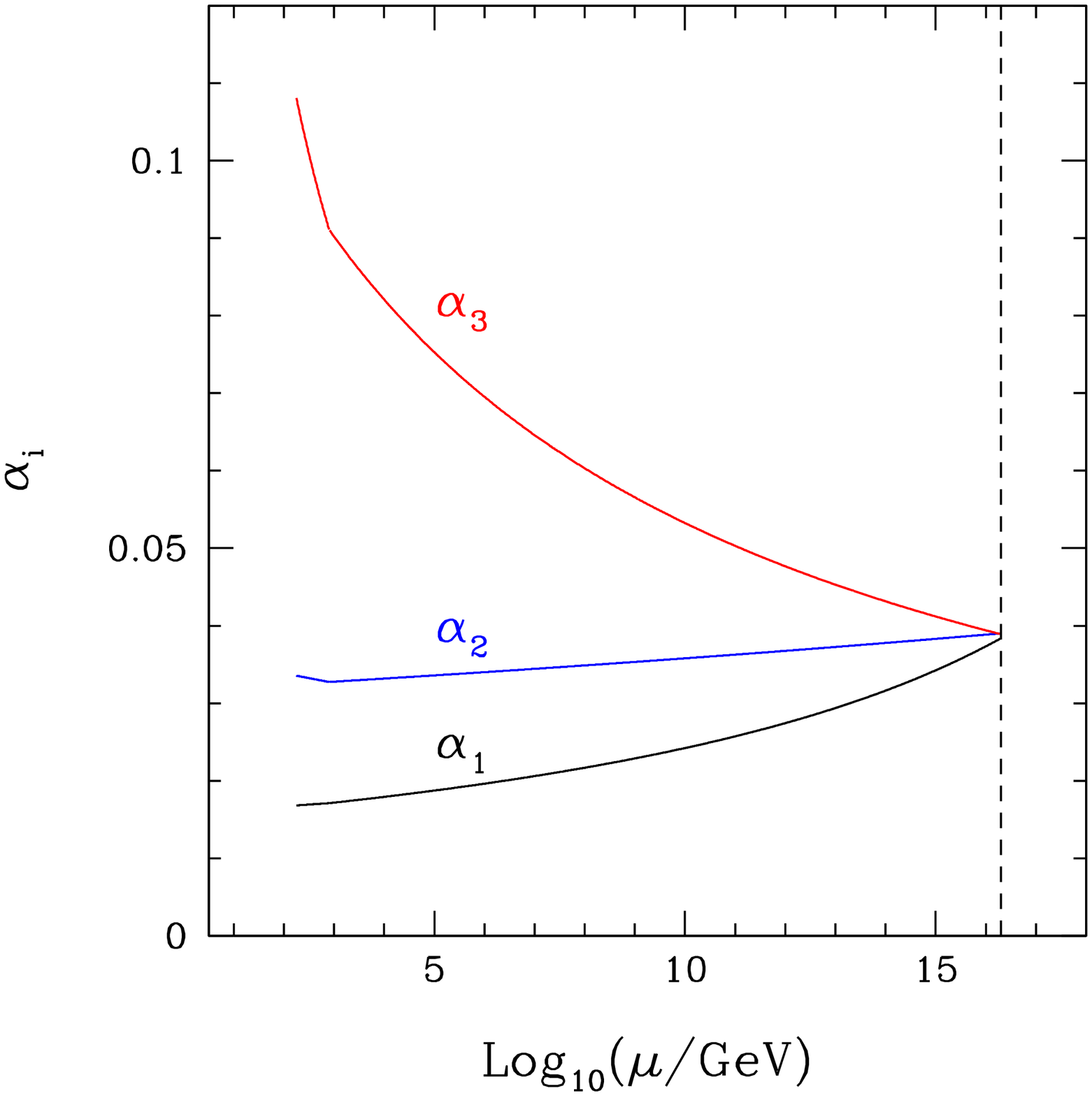}
\includegraphics[height=7cm,width=7cm]{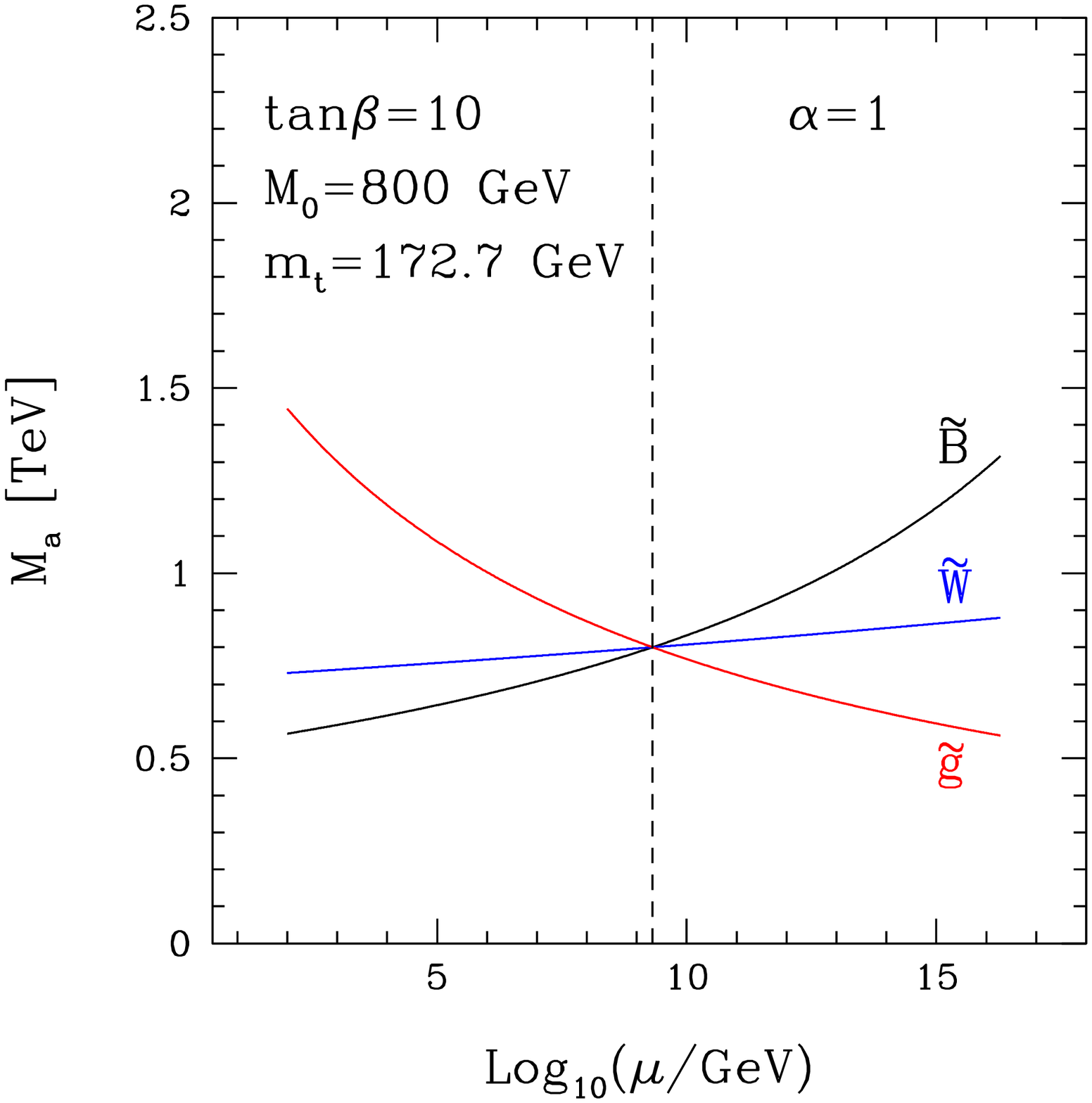}
\includegraphics[height=7cm,width=7cm]{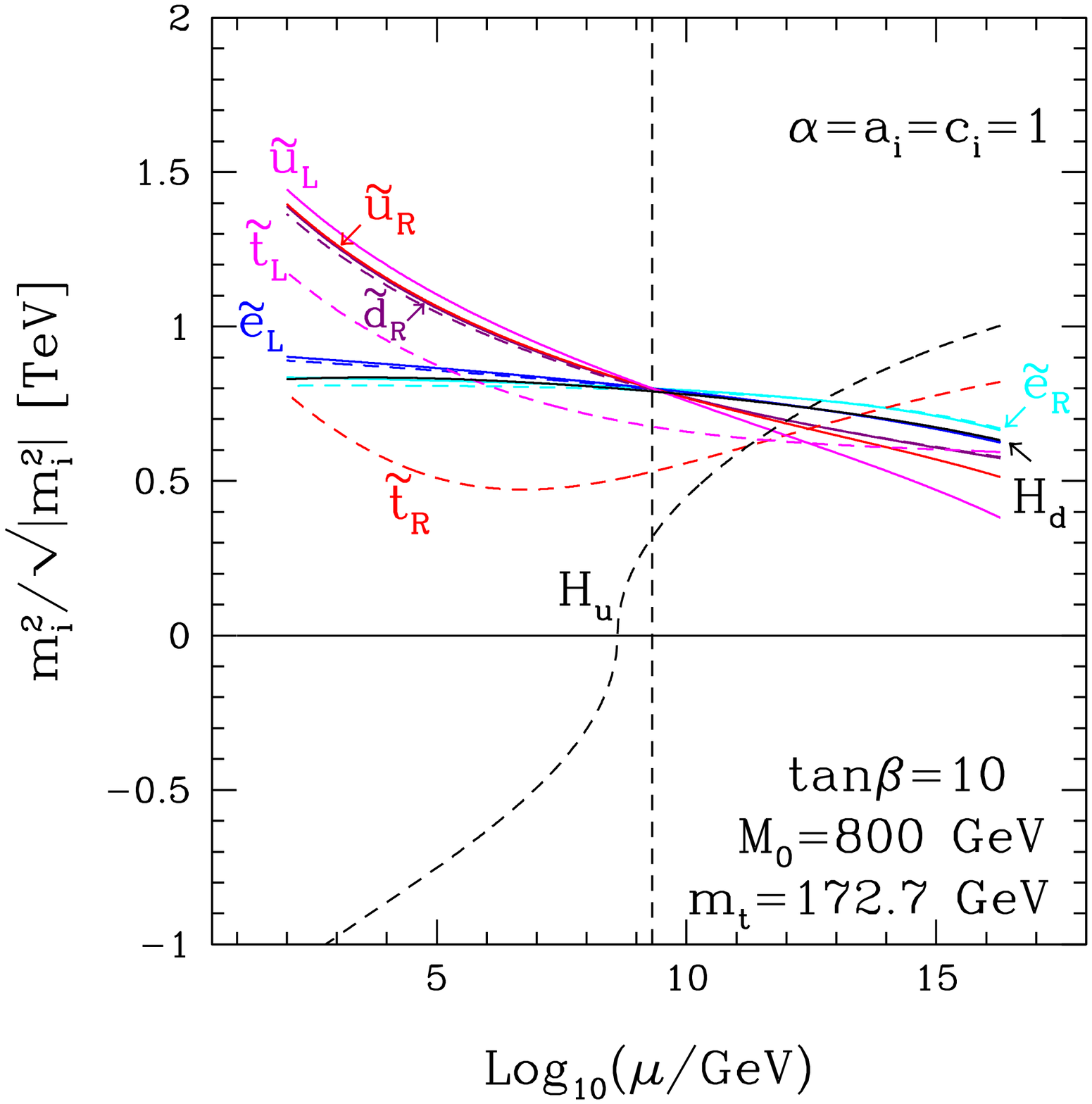}
\includegraphics[height=7cm,width=7cm]{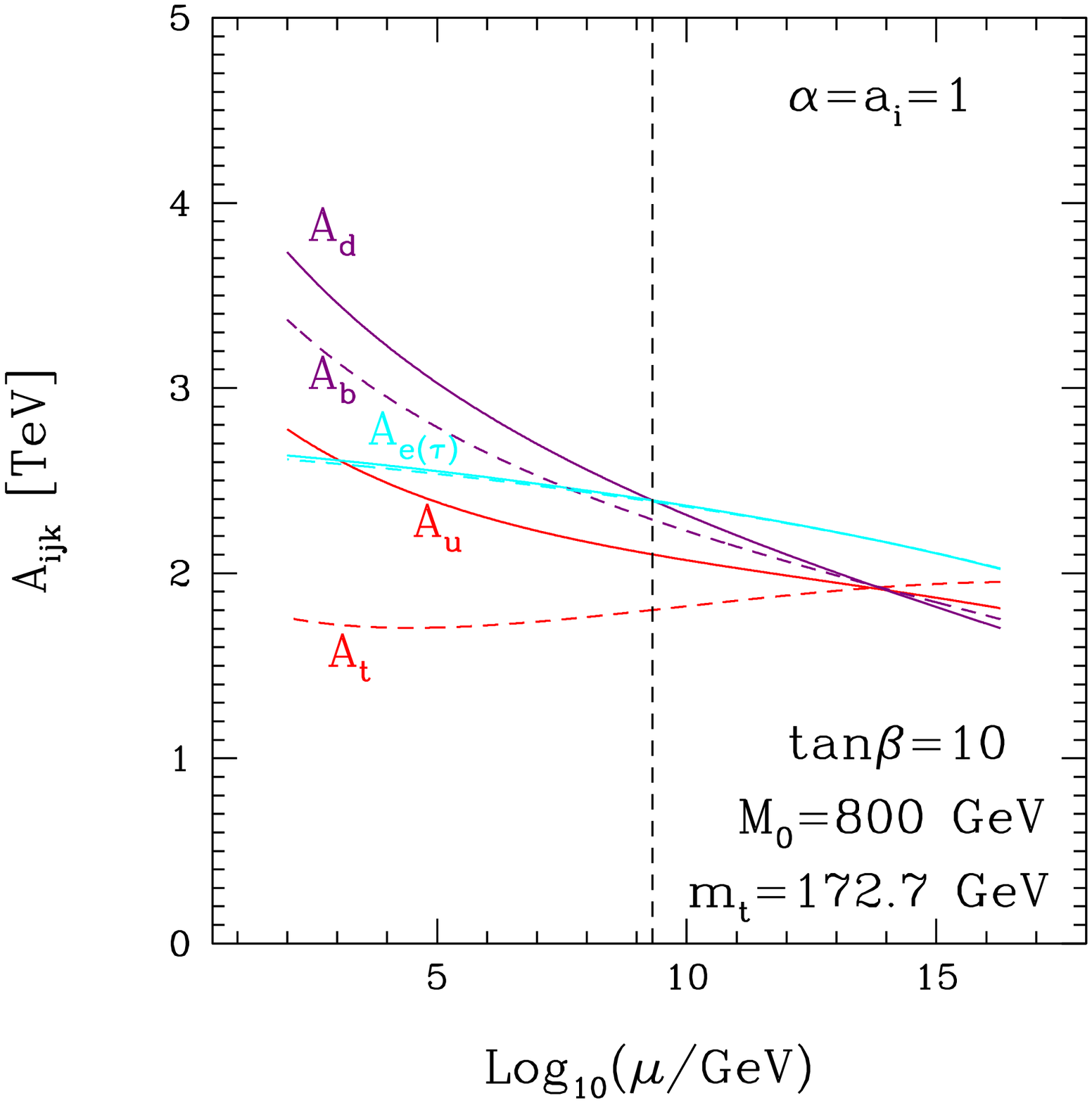}
\end{center}
\vskip -0.5cm \caption{RG evolution of (a) gauge couplings
$\alpha_i$, (b) gaugino masses $M_a$, (c) sfermion and Higgs masses
$m_i$, (d) trilinear $A$ parameters in intermediate scale mirage
mediation with $a_i=c_i=1$. Here we choose $M_0=800$ GeV and
$\tan\beta=10$.} \label{fig:rge}
\end{figure}

In  KKLT compactification of type IIB string theory \cite{kklt},
$T$ corresponds to the Calabi-Yau  volume modulus, and the
uplifting brane is located at the end of warped throat. In this
case, the uplifting operator is sequestered from $T$ \cite{choi1}:
\bea \label{sequestered}\partial_T {\cal P}_{\rm lift}=0. \eea
Also, the minimal  KKLT compactification of IIB theory gives \bea
\label{minimal} K_0&=&-3\ln(T+T^*), \quad
Z_i\,=\,\frac{1}{(T+T^*)^{n_i}}, \nonumber \\
f_a&=& kT, \quad W_0=w-Ae^{-aT} \quad (A\,=\, {\cal O}(1)),\eea
where the modular weight $n_i$ is a rational number depending on
the origin of matter superfield $\Phi_i$. Then from
(\ref{tmediation}) and (\ref{vev}), one immediately finds \bea
\label{benchmark} \alpha\,=\, 1,\quad a_i\,=\, c_i\,=\, 1-n_i,
\eea giving an intermediate mirage messenger scale: \bea M_{\rm
mir}\,\sim\, 3\times 10^9 \,\, \mbox{GeV}. \eea
 If $\Phi_i$ lives on the entire world-volume of $D7$ brane
from which the visible gauge bosons originate, the corresponding
modular weight $n_i=0$. However, if $\Phi_i$ is confined in the
intersections of $D7$ branes, $n_i$ has a positive value, e.g.
$n_i=1/2$ or 1. In Fig.~\ref{fig:rge}, we depict the RG evolution
of gauge couplings and soft masses in intermediate scale mirage
mediation scenario with $n_i=0$, i.e.  $\alpha=a_i=c_i=1$, which
shows that indeed the gaugino masses and the 1st and 2nd
generations of squarks and slepton masses are unified at $M_{\rm
mir}\sim 3\times 10^9$ GeV as indicated by (\ref{lowgaugino}) and
(\ref{lowsfermion}). Note that still the gauge couplings are
unified at the conventional GUT scale $M_{GUT}\sim 2\times
10^{16}$ GeV. In view of its minimality, intermediate scale mirage
mediation can be considered as a benchmark scenario, thus we
perform a detailed analysis of neutralino DM in intermediate scale
mirage mediation in the next section.

It is in fact easy to generalize the compactification to get
different values of the mirage mediation parameters $\alpha$,
$a_i$ and $c_i$. For instance, the compactification  can be
generalized to have a dilaton-modulus mixing in gauge kinetic
functions \cite{ck,lust} and/or the non-perturbative
superpotential \cite{abe}. For the case of type IIB
compactification, a nonzero gauge flux on $D7$ branes can generate
such dilaton-modulus mixing in $D7$ gauge kinetic functions
\cite{lust}, which would result in \bea f_a\,=\,kT+lS_0, \quad
W_0\,=\,w-Ae^{-(aT+bS_0)}\quad (A={\cal O}(1)),\eea where $S_0$
denotes the vacuum value of the string dilaton $S$ which is
assumed to get superheavy mass from RR and NS-NS 3-form fluxes,
and $k,l,a$ and $b$ are real parameters. In such IIB
compactifications, the coefficients of $T$ in $D7$ gauge kinetic
functions and non-perturbative superpotential, i.e. $k$ and $a$,
are positive, however the coefficients of $S$, i.e. $l$ and $b$,
might have both signs under the conditions:\bea
\label{condition}k{\rm Re}(T)+l{\rm Re}(S_0)&\simeq&
\frac{1}{g_{GUT}^2}\,\simeq\, 2,
\nonumber \\
a{\rm Re}(T)+b{\rm Re}(S_0)&\simeq& \ln(M_{Pl}/m_{3/2})\,\simeq\,
4\pi^2.\eea Assuming that $e^{-K_0/3}Z_i$ has the same form as the
minimal model (\ref{minimal}), it is then straightforward to find
that the mirage mediation
parameters are given by \bea\label{general} \alpha&=&\frac{1+R_1}{(1+R_2)(1+R_3)},\nonumber \\
a_i&=& (1-n_i)(1+R_1),\nonumber \\
c_i&=& (1-n_i)(1+R_1)^2 \eea where \bea R_1=\frac{l{\rm
Re}(S_0)}{k{\rm Re}(T)}, \quad R_2=\frac{b{\rm Re}(S_0)}{a{\rm
Re}(T)},\quad R_3=\frac{3\partial_T\ln({\cal P}_{\rm
lift})}{2\partial_T K_0}.\eea
 Again, for ${\cal P}_{\rm lift}$
induced by an uplifting brane at the end of warped throat,
$R_3=0$. However, $1+R_{1}$ and $1+R_2$ can have a variety of
(positive) values under the condition (\ref{condition}). As a
result, the anomaly to modulus mediation ratio $\alpha$ can easily
have any (positive) value within the range of order unity.


\begin{figure}[ht!]
\begin{center}
\includegraphics[height=7cm,width=7cm]{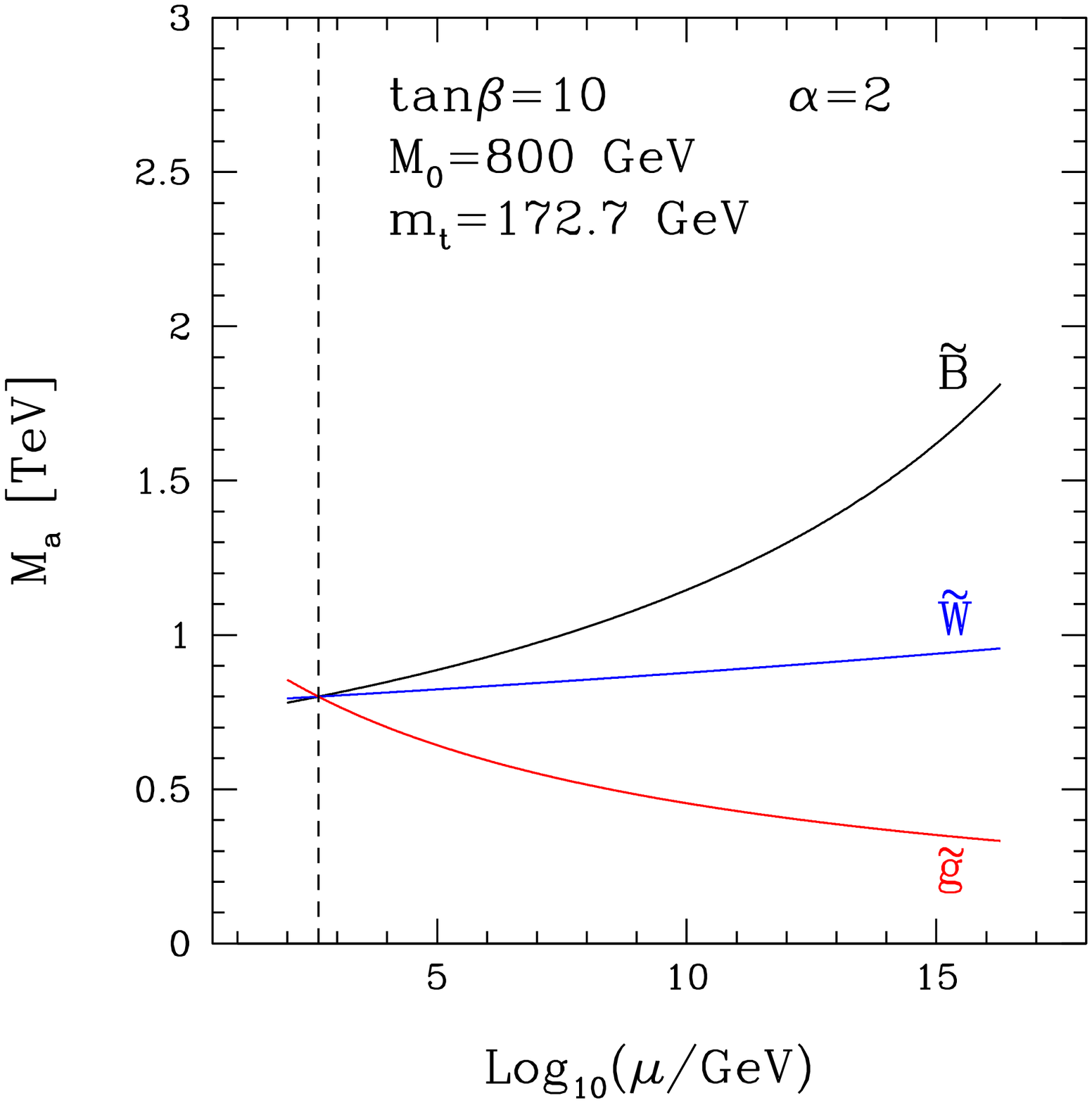}
\includegraphics[height=7cm,width=7cm]{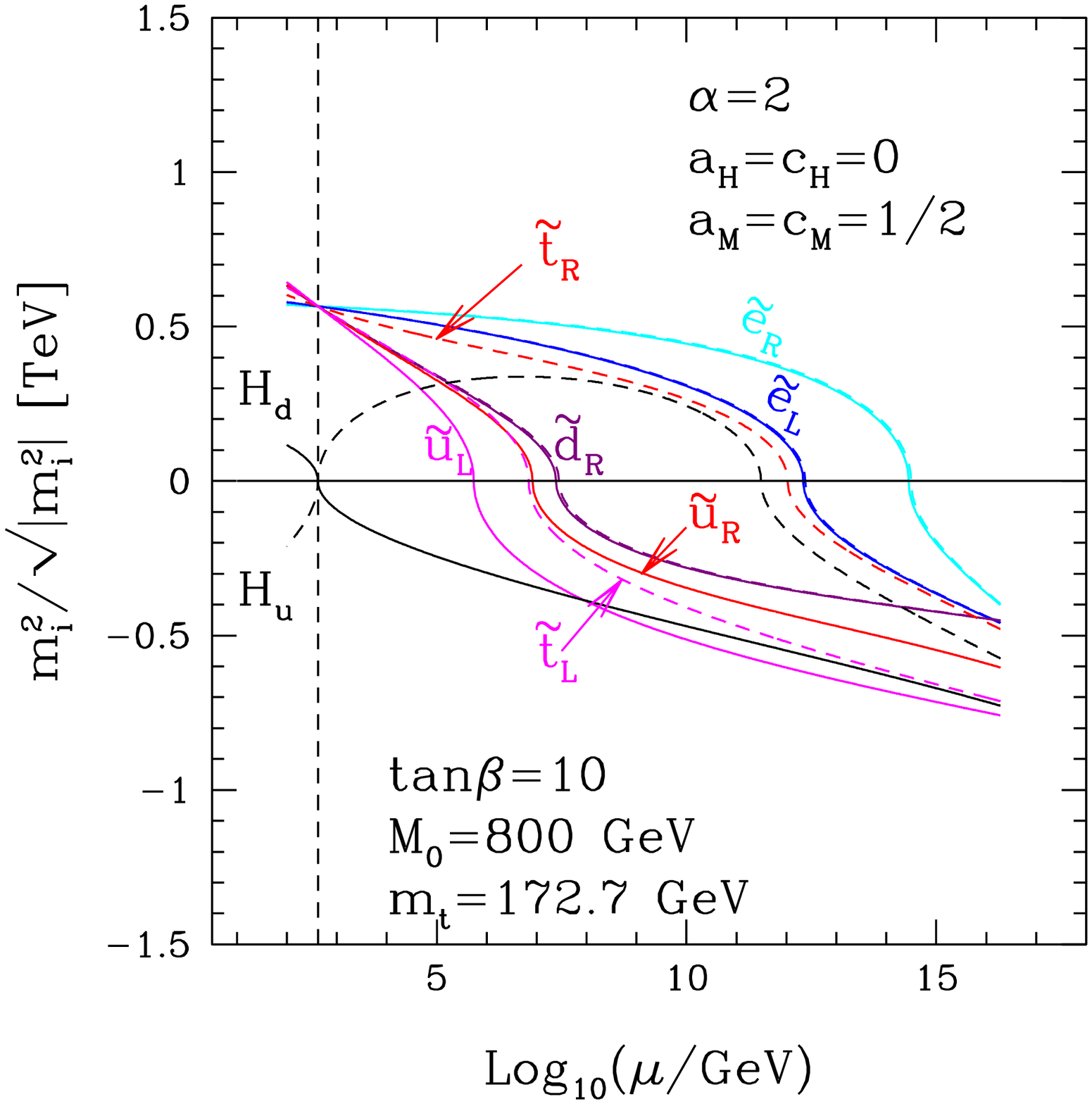}
\end{center}
\vskip -0.5cm \caption{RG evolution of (a) gaugino masses and (b)
sfermion and Higgs masses in TeV scale mirage mediation. Here we
fixed $M_0=800$ GeV, $a_{\rm H}=c_{\rm H}=0$ and $a_{\rm M}=c_{\rm
M}=1/2$.} \label{fig:rge2}
\end{figure}

A particularly interesting model is the TeV scale mirage mediation
with \bea \alpha\,=\,2,\quad a_{H_u}\,=\, c_{H_u}\,=\, 0, \quad
a_{Q_3}+a_{U_3}=c_{Q_3}+c_{U_3}=1, \eea where
$Q_3$ and $U_3$ denote the left-handed and right handed top-quark
superfields, respectively. This particular model has been claimed
to minimize the fine tuning for the electroweak symmetry breaking
in the MSSM \cite{tevmirage}. In Fig.~\ref{fig:rge2}, we depict
the RG evolution of soft masses in TeV scale mirage mediation
model with $a_{\rm H}=c_{\rm H}=0$ and $a_{\rm M}=c_{\rm M}=1/2$,
where the subscripts H and  M stands for the MSSM Higgs doublets
$H_{u,d}$ and the quark/lepton matter superfields, respectively.

Note that the squark/slepton mass-squares renormalized  at high
energy scale, e.g. at a scale near $M_{GUT}$,  are negative in
this model, while the values at low energy scale below $10^6$ GeV
are positive.
 In fact,
such tachyonic {\it high energy} squark/slepton mass-squares is a
generic feature of mirage mediation for $\alpha>\alpha_c$ where
the precise value of $\alpha_c$ depends on $a_i$ and $c_i$, but
not significantly bigger than 1 in most cases. As long as the low
energy squark/slepton mass-squares are positive, the model has a
correct color/charge preserving (but electroweak symmetry
breaking) vacuum. For instance, the TeV scale mirage mediation
model of Fig.~\ref{fig:rge2} has a such vacuum which is a local
minimum of the scalar potential over the squark/slepton values
$|\phi|\lesssim 10^6$ GeV. On the other hand, tachyonic  squark
mass-squares at the RG point $\mu > 10^6$ GeV indicates that there
might be a deeper CCB minimum color/charge breaking (CCB) or an
unbounded from below (UFB) direction at $|\phi|> 10^6$ GeV. One
then needs a cosmological scenario which allows our universe to be
settled down at the correct vacuum with $\phi=0$. In view of that
the squarks and sleptons get large positive mass-squares in the
high temperature limit, it is rather plausible assumption that
squark/sleptons are settled down at the color/charge preserving
minimum after the inflation \cite{kuzenko}. One still needs to
confirm that the color/charge preserving vacuum is stable enough
against the decay into CCB vacuum. It has been noticed that the
corresponding tunnelling rate is  small enough, i.e. less than the
Hubble expansion rate, as long as the RG points of vanishing
squark/slepton mass-squares are all higher than $10^4$ GeV
\cite{riotto,kuzenko}, which is satisfied safely by the TeV scale
mirage mediation of Fig.~\ref{fig:rge2}.

\section{Neutralino DM in intermediate scale mirage mediation}


In this section, we examine the prospect of neutralino DM in
intermediate scale mirage mediation scenario. As was noticed in
the previous section, the minimal KKLT-type model (\ref{minimal})
with a sequestered uplifting brane  gives $\alpha=1$, thus an
intermediate mirage messenger scale \bea M_{\rm mir}\sim
M_{GUT}(m_{3/2}/M_{Pl})^{1/2}\sim 3\times 10^9\,\, {\rm GeV}.\eea
In this minimal set-up, the discrete parameters $a_i$ and $c_i$
describing the modulus mediated $A$-parameters and sfermion masses
are determined to be $a_i=c_i=1-n_i$, where $n_i$ denote the
modular weights of matter and Higgs superfields. Throughout this
paper, we will assume $a_i=c_i$ and consider the following four
different cases: \bea \label{4choices} (a_{\rm H}=c_{\rm H},\,
a_{\rm M}=c_{\rm M})=(1,\,1), \quad (\frac{1}{2},\,\frac{1}{2}),
\quad (0,\,1), \quad (0,\,\frac{1}{2}). \eea We also choose the
Higgsino mass parameter $\mu > 0$ in light of the experimental
value of the muon anomalous magnetic moment which favors positive
$\mu$ \cite{g-2}, and treat $\tan\beta=\langle H_u\rangle/\langle
H_d\rangle$ as a free parameter without specifying the origin of
the corresponding $\mu$ and $B$ parameters. We then obtain the
parameter range of the model for which the LSP is the lightest
neutralino as well as the relic neutralino DM abundance under the
assumption of thermal production, and finally  the direct and
indirect detection rates of the neutralino LSP using the DarkSUSY
routine \cite{darksusy}.

\begin{figure}[ht!]
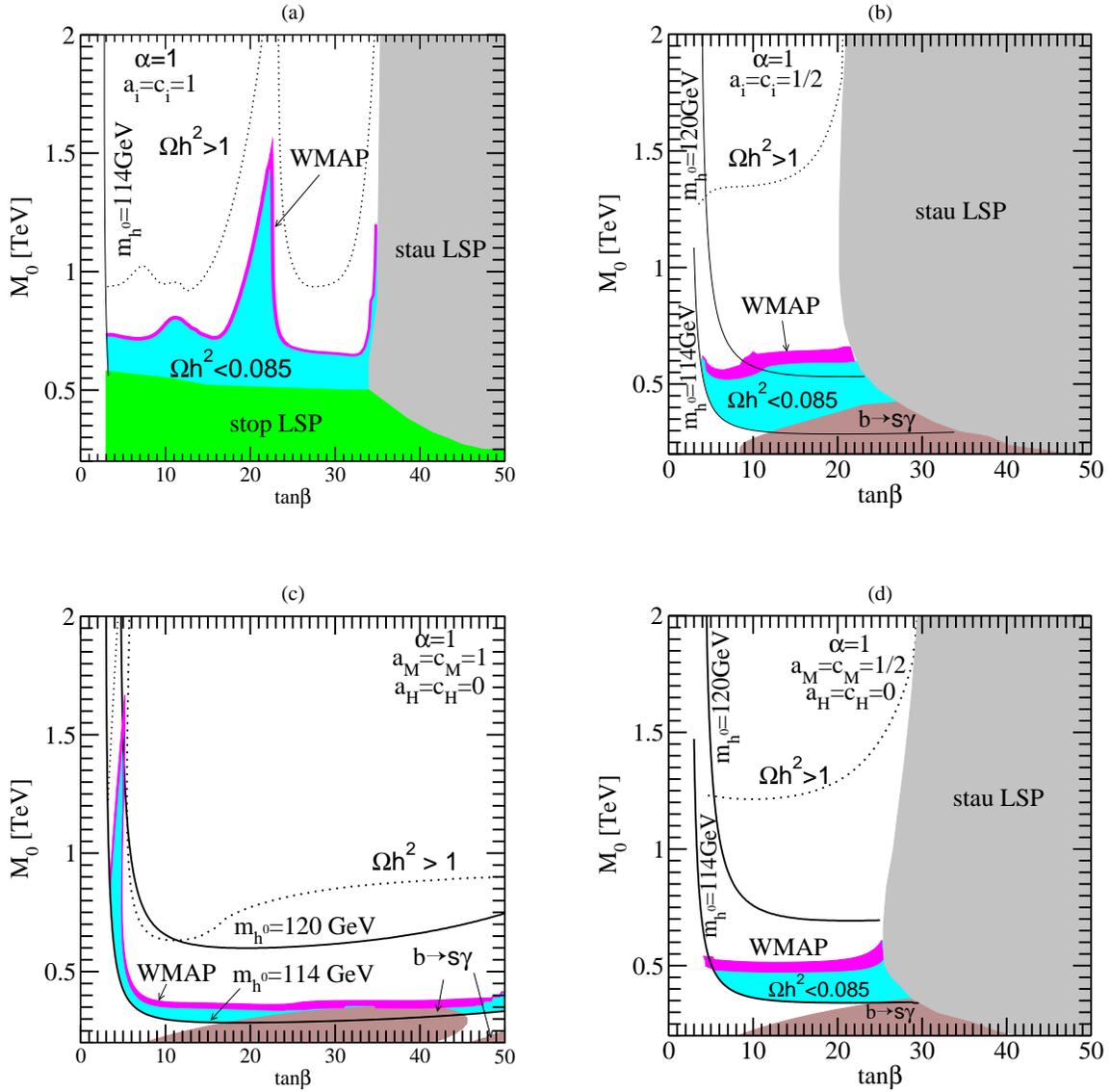

\begin{center}
\includegraphics[height=7cm,width=7cm]{t-m0.eps}
\hskip 1cm
\includegraphics[height=7cm,width=7cm]{t-m0-nq05-nh05.eps}
\vskip 1cm
\includegraphics[height=7cm,width=7cm]{t-m0-nq0-nh1.eps}
\hskip 1cm
\includegraphics[height=7cm,width=7cm]{t-m0-nq05-nh1.eps}
\end{center}
\vskip -0.4cm \caption{ Parameter region of neutralino LSP and the
thermal relic density depicted on the plane of $(\tan\beta, M_0)$
in intermediate scale mirage mediation models with $(a_i,c_i)$
specified in Eq.~(3.2).
}
\label{fig:density}
\end{figure}

\subsection{Parameter region of neutralino LSP and thermal relic density}

In Fig.~\ref{fig:density}, we show the neutralino LSP region and
the thermal neutralino relic density in intermediate scale mirage
mediation scenario on the ($\rm{tan}\beta$,$M_0$)-plane for the
values of
 $a_i$ and $c_i$ specified in Eq.~(\ref{4choices}). 
We computed  the sparticle mass spectrum at the electroweak  scale
by solving the RG equations with the boundary condition
(\ref{soft1}) at $M_{GUT}$. Our results show that in all cases
there is a large parameter region for which the LSP is given by
the lightest neutralino.

\begin{figure}[ht!]
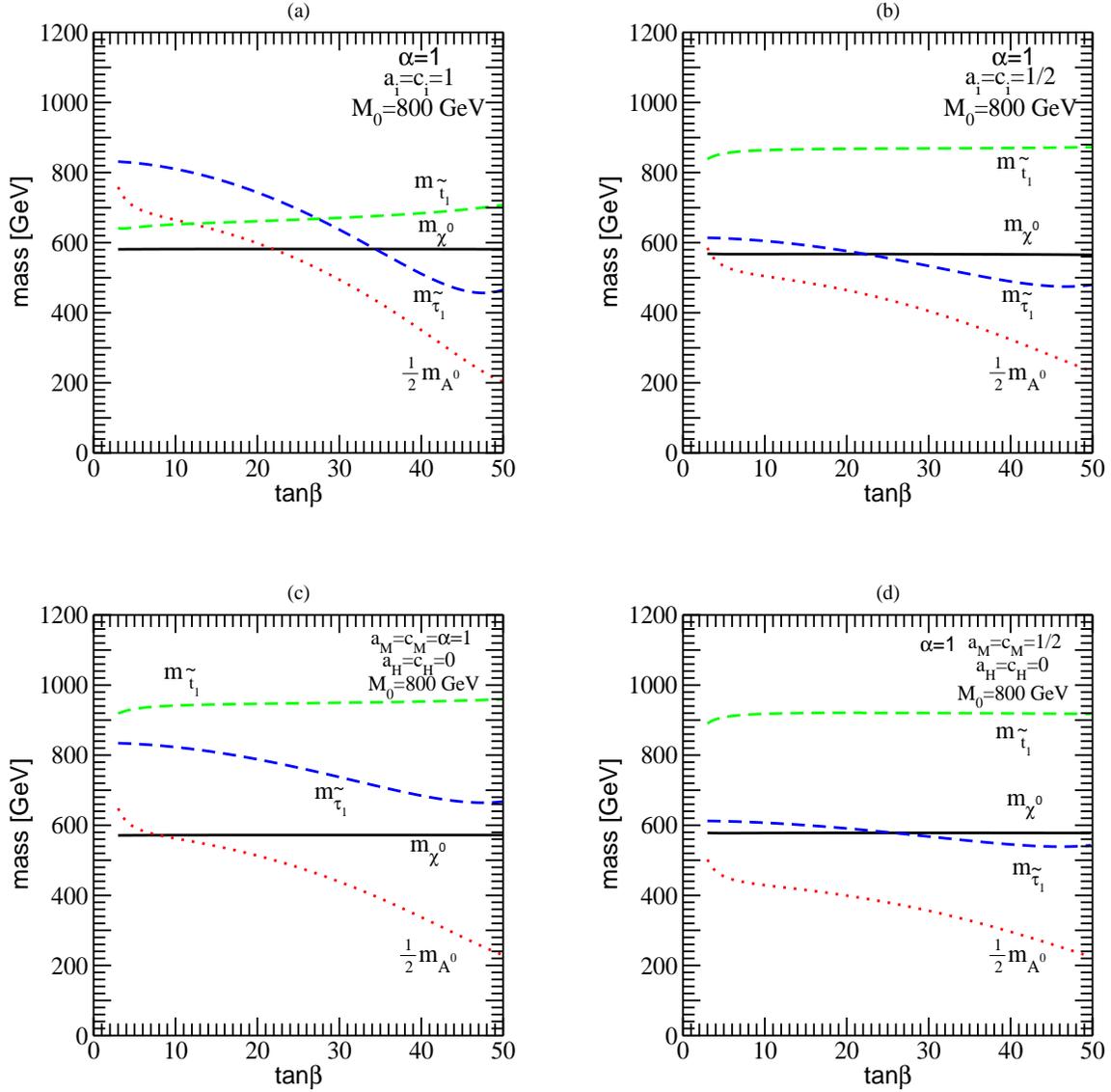

\begin{center}
\includegraphics[height=7cm,width=7cm]{mass.eps}
\hskip 1cm
\includegraphics[height=7cm,width=7cm]{nzmw-alpha1-mass.eps}
\vskip 1cm
\includegraphics[height=7cm,width=7cm]{nzmw2-alpha1-mass.eps}
\hskip 1cm
\includegraphics[height=7cm,width=7cm]{nzmw3-alpha1-mass.eps}
\end{center}
\vskip -0.5cm \caption{ Particle masses as a function of $\tan
\beta$ in intermediate scale mirage mediation models with
$(a_i,c_i)$ specified in Eq.~(3.2).
Here, we fixed $M_0=800$ GeV.}
\label{fig:higgs}
\end{figure}

Fig.~\ref{fig:density}.a is the result for the case in which
$a_i=c_i=1$ for both matter and Higgs multiplets. In this case,
large $\rm{tan}\beta >34$ (grey color) for which the tau Yukawa
couplings becomes sizable gives stau LSP (see also Fig.
\ref{fig:higgs}.a), while small $M_0\lesssim0.5~ \rm TeV$ (green color)
gives stop LSP. In the remaining region, the LSP is the lightest
neutralino which turns out to be Bino-like. Small
$\tan\beta\lesssim 3$ is excluded by the Higgs mass limit $m_h >
114$ GeV.
Under the assumption that the DM neutralinos are produced purely
by the conventional thermal production mechanism, the magenta stripe corresponds to
the parameter region giving a relic neutralino density consistent with   the recent WMAP observation
\cite{wmap}:
\begin{eqnarray}
0.085 < \Omega_{\rm DM} h^2 < 0.119 ~~(2\sigma~ \rm level).
\label{eq:wmap}
\end{eqnarray}
In the region below the magenta stripe, $\Omega_\chi h^2 < 0.085$,
while $\Omega_\chi h^2 > 0.119$ for the upper region. Thus the
(cyan) region below the magenta stripe (but above the stop LSP
region) can be phenomenologically viable if additional DM
neutralinos were produced by non-thermal mechanism such as the
decays of flaton in thermal inflation \cite{stewart}.

\begin{figure}[ht!]
\begin{center}
\includegraphics[height=7cm,width=7cm]{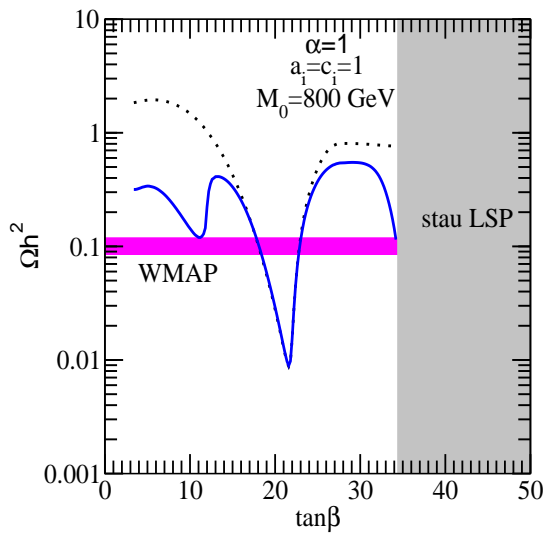}
\end{center}
\vskip -0.5cm \caption{$\Omega_\chi h^2$ as a function of
$\tan\beta$ in intermediate scale mirage mediation with
$a_i=c_i=1$ and $M_0=800$ GeV. Here the dotted line corresponds to
the result computed without including the stop/stau coannihilation
effects.} \label{fig:coan}
\end{figure}

Although the neutralino LSP  is Bino-like in this particular
intermediate scale mirage mediation, the WMAP mass density is
obtained for a rather heavy neutralino mass $m_{\chi^0}\gtrsim
450$ GeV. This can be understood by Fig.~\ref{fig:higgs}.a which
shows the masses of the lightest neutralino, lighter stop and
stau, and also the pseudo-scalar Higgs boson as a function of
$\tan\beta$ for the model with $\alpha=1$, $a_i=c_i=1$ and
$M_0=800$ GeV. Around $\tan\beta\sim 20$, the pseudoscalar Higgs
mass  becomes same as $2m_{\chi^0}$, leading to a resonant
enhancement of neutralino annihilation through the s-channel
pseudo-scalar Higgs exchange. For other values of $\tan\beta$, the
neutralino mass is somewhat close to the stop mass  (or to the
stau mass at $\tan\beta\sim 34$), making the stop-neutralino  or
stau-neutralino coannihilation process becomes efficient. In
Fig.~\ref{fig:coan}, we depicted $\Omega_\chi h^2$  as a function
of $\rm tan\beta$ for $M_0 = 800$ GeV, which shows clearly the
effect of Higgs resonance at $\tan\beta\sim 20$ and also the
effect of stop/stau coannihilation effects for other values of
$\tan\beta$. On the plot, the dotted line corresponds to the relic
density computed without including coannihilation effects. It
indicates that the stop/stau-neutralino coannihilation plays a
crucial role for the relic neutralino density to have the WMAP
value (\ref{eq:wmap}) for $M_0=700\sim 800$ GeV and $\tan\beta$
outside the Higgs resonance region.  Note that in intermediate
scale mirage mediation with $a_i=c_i=1$, the pseudoscalar Higgs
resonance condition $m_A \simeq 2 m_\chi$ is satisfied for smaller
value of $\rm tan\beta$ compared to the mSUGRA case. This can be
understood by noting that the low energy gaugino masses in mirage
mediation are more compressed compared to mSUGRA, e.g. $M_3 /M_1
\sim 2.3$ in the intermediate scale mirage mediation with $\alpha
= 1$, while $M_3 /M_1 \sim 6$ in mSUGRA. For a given value of
$M_1$, smaller $M_3$ gives smaller $\mu$ and $m_A^2 \sim m_{H_d}^2
+ \mu^2$ at the electroweak scale, thus the pseudoscalar Higgs
resonance appears at smaller value of $\rm tan\beta$ compared to
mSUGRA case.

\vskip 0.8cm
%

So far, we have been focusing on the specific intermediate scale
mirage mediation model with $a_i=c_i=1$ which might be obtained
when all modular weights $n_i=0$. However as anticipated in the
previous section, $a_i=c_i=1$ is not necessarily a more favored
choice than the other values of $(a_i,c_i)$ in
Eq.~(\ref{4choices}). Different values of $a_i$ and $c_i$, e.g.
smaller but still non-negative values, are also equally plausible.
Obviously, for a fixed value of $M_0$, the gaugino masses are not
affected by changing $a_i$ and $c_i$.
However the low energy stop, stau and Higgs masses are somewhat
sensitive to the values of $a_i$ and $c_i$. They depend on $a_i$
and $c_i$ either through their boundary values at $M_{GUT}$, or
through their RG evolutions, or through the mass-mixing induced by
the low energy $A$-parameters.

The effects of changing $a_i$ and $c_i$ on the RG evolution can be
read off from the following one-loop RG equations for the Higgs
and third generation sfermion mass-squares:
\begin{eqnarray}
\label{rgequation} 16\pi^2{d\over dt} m_{H_u}^2 &=& 3 X_t -6 g_2^2
|M_2|^2
- {6\over 5} g_1^2 |M_1|^2 ,\nonumber \\
16\pi^2{d\over dt} m_{H_d}^2 &=& 3 X_b + X_\tau-6 g_2^2 |M_2|^2
- {6\over 5} g_1^2 |M_1|^2 ,\nonumber \\
16\pi^2{d\over dt} m_{Q_3}^2 &=& X_t+X_b-{32\over 3} g_3^2 |M_3|^2
-6 g_2^2 |M_2|^2-{2\over 15} g_1^2 |M_1|^2, \nonumber \\
16\pi^2{d\over dt} m_{U_3}^2 &=& 2 X_t-{32\over 3} g_3^2 |M_3|^2
-{32\over 15} g_1^2 |M_1|^2, \nonumber \\
16\pi^2{d\over dt} m_{L_3}^2 &=& X_\tau
-6 g_2^2 |M_2|^2-{3\over 5} g_1^2 |M_1|^2, \nonumber \\
16\pi^2{d\over dt} m_{E_3}^2 &=& 2 X_\tau -{24\over 5} g_1^2
|M_1|^2,
\end{eqnarray}
where
\begin{eqnarray}
X_t &=& 2 y_t^2 (m_{H_u}^2+m_{Q_3}^2+m_{U_3}^2+ A_{H_uQ_3U_3}^2), \nonumber \\
X_b &=& 2 y_b^2 (m_{H_d}^2+m_{Q_3}^2+m_{D_3}^2+ A_{H_dQ_3D_3}^2), \nonumber \\
X_\tau &=& 2 y_\tau^2 (m_{H_d}^2+m_{L_3}^2+m_{E_3}^2+
A_{H_dL_3E_3}^2).
\end{eqnarray}
These RG equations show that smaller $X_I$ ($I=t,b,\tau$) increase
the low energy soft mass-squares.
 Since $a_i$ and $c_i$ determine
 the modulus-mediated trilinear $A$
parameters and soft mass-squares at $M_{GUT}$ as
$\tilde{A}_{ijk}=(a_i+a_j+a_k)M_0$ and $\tilde{m}_i^2=c_iM_0^2$,
smaller $a_{\rm H}=c_{\rm H}$ give smaller $X_I$  without
affecting the boundary values of squark and slepton masses at
$M_{GUT}$, eventually making the stop and stau masses at TeV scale
larger. On the other hand, the consequence of smaller $a_{\rm
M}=c_{\rm M}$ is more complicate as it depends on the relative
importance of the Yukawa-induced RG evolution. It turns out that
changing $a_{\rm M}=c_{\rm M}$ to smaller value makes the stop
mass larger, while the stau mass smaller.

In Figs.~\ref{fig:density}.b and \ref{fig:higgs}.b, we depict the
results for the case in which  $a_i=c_i=1/2$ for both the matter
and Higgs multiplets. As can be understood from the above
discussion, this intermediate scale mirage mediation
does not contain any parameter region of stop LSP, while having a
larger parameter region of stau LSP (see Fig.~\ref{fig:higgs}.b).
Another important feature is that the weak scale value of
$|\,m_{H_u}^2|$  becomes smaller compared to the case of
$a_i=c_i=1$, which is mainly due to smaller $X_t$.
 This results in smaller $\mu$ and $m_A$. Smaller $\mu$ makes the
neutralino LSP have a sizable Higgsino component, while smaller
$m_A$ makes the pseudo-scalar resonance region disappear. Again
the magenta region in Fig. \ref{fig:density}.b corresponds to the
parameter region giving the WMAP DM density (\ref{eq:wmap}) under
the assumption of pure thermal production. In this case, the
neutralino pair annihilation into gauge boson pair becomes
efficient due to the enhanced Higgsino component of neutralino
LSP. Finally, the brown region is excluded by giving the Br($b \to
s \gamma$) smaller than the allowed range.

Figs.~\ref{fig:density}.c and \ref{fig:higgs}.c are the result for
the case with $a_{\rm M}=c_{\rm M}=1$ and $a_{\rm H}=c_{\rm H}=0$,
while Figs.~\ref{fig:density}.d and \ref{fig:higgs}.d are for the
case with $a_{\rm M}=c_{\rm M}=1/2$ and $a_{\rm H}=c_{\rm H}=0$.
The case of $a_{\rm M}=c_{\rm M}=1/2$ and $a_{\rm H}=c_{\rm H}=0$
is quite similar to the case of $a_i=c_i=1/2$: LSP is the lightest
neutralino with a sizable Higgsino component for
$\tan\beta\lesssim 20$ (see Figs.~\ref{fig:higgs}.b and
\ref{fig:higgs}.d). On the other hand, the case of $a_{\rm
M}=c_{\rm M}=1$ and $a_{\rm H}=c_{\rm H}=0$ is somewhat
distinctive since there is no parameter region of stop or stau LSP
and the WMAP DM density is obtained for a light neutralino mass
$m_{\chi^0}\sim 250$ GeV, while in other cases the WMAP DM density
is obtained for heavier $m_{\chi^0}\gtrsim 350$ GeV. Again, the
brown region is excluded by giving the Br($b \to s \gamma$)
smaller than the allowed range.

\subsection{Dark matter detections}

\begin{figure}[ht!]
\begin{center}
\includegraphics[height=7cm,width=7cm]{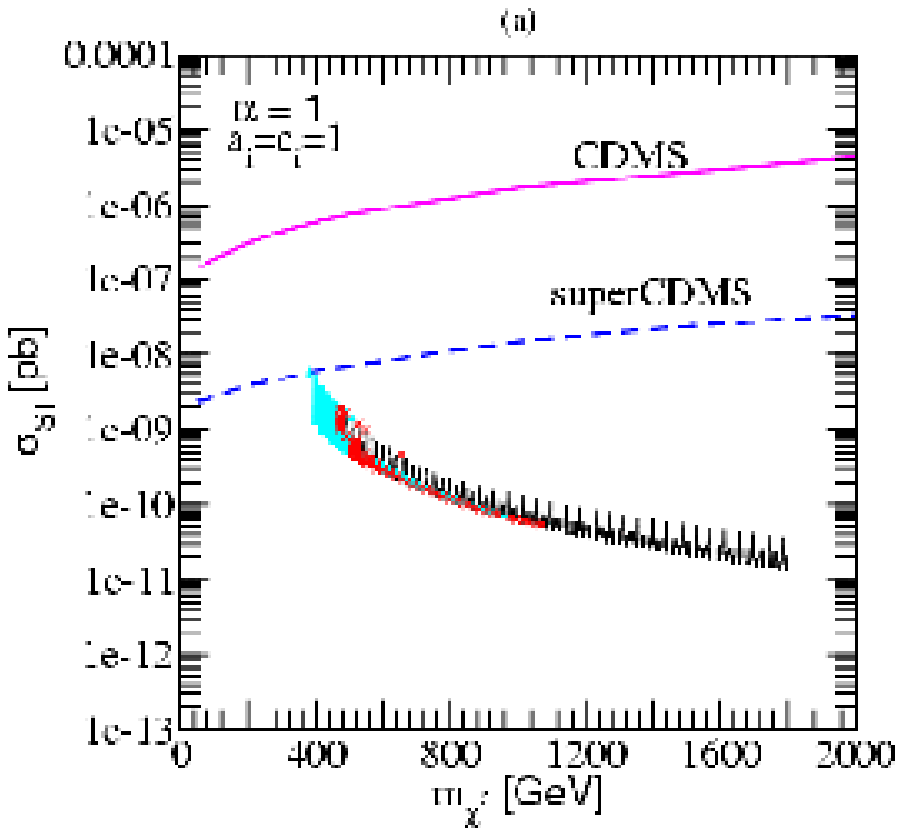}
\hskip 1cm
\includegraphics[height=7cm,width=7cm]{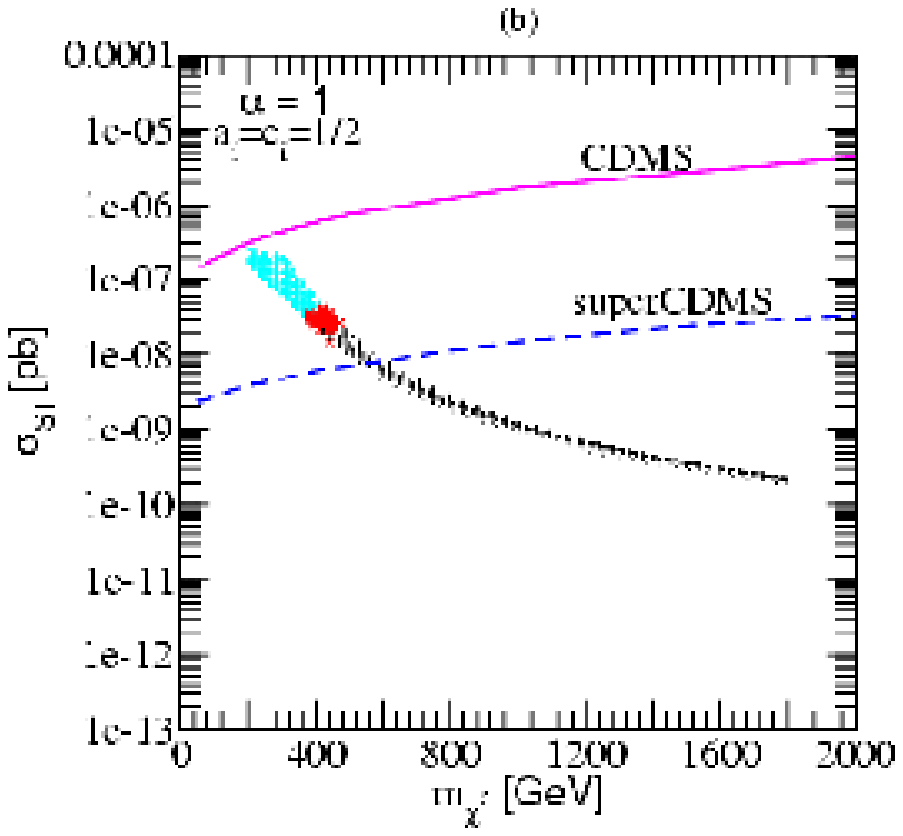}
\vskip 1.2cm
\includegraphics[height=7cm,width=7cm]{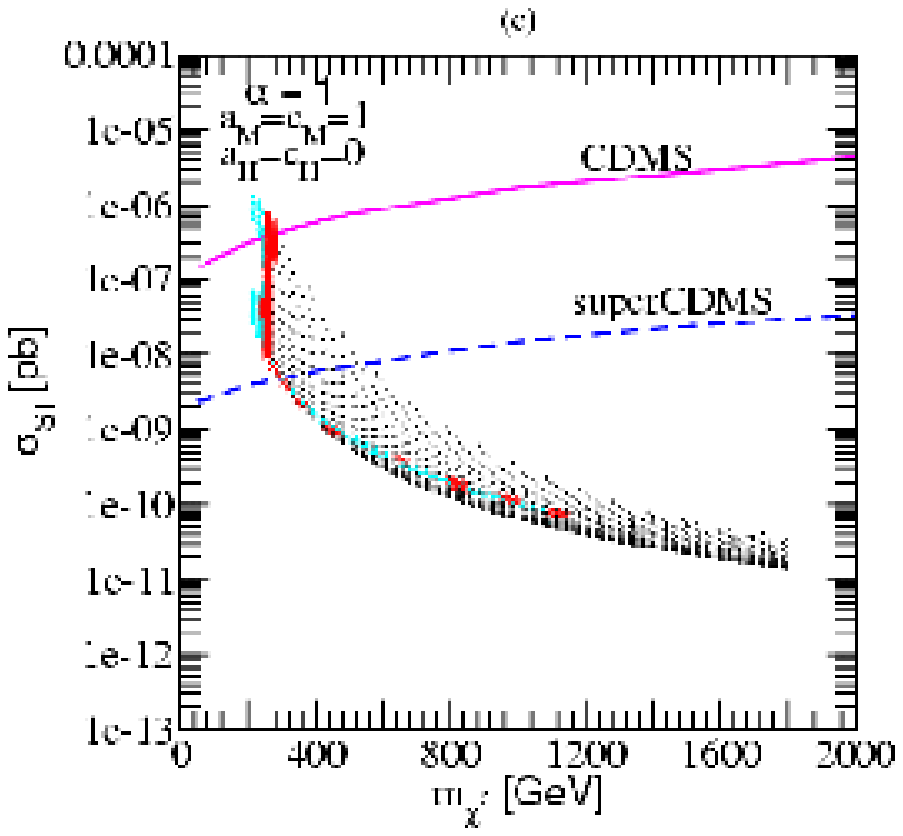}
\hskip 1cm
\includegraphics[height=7cm,width=7cm]{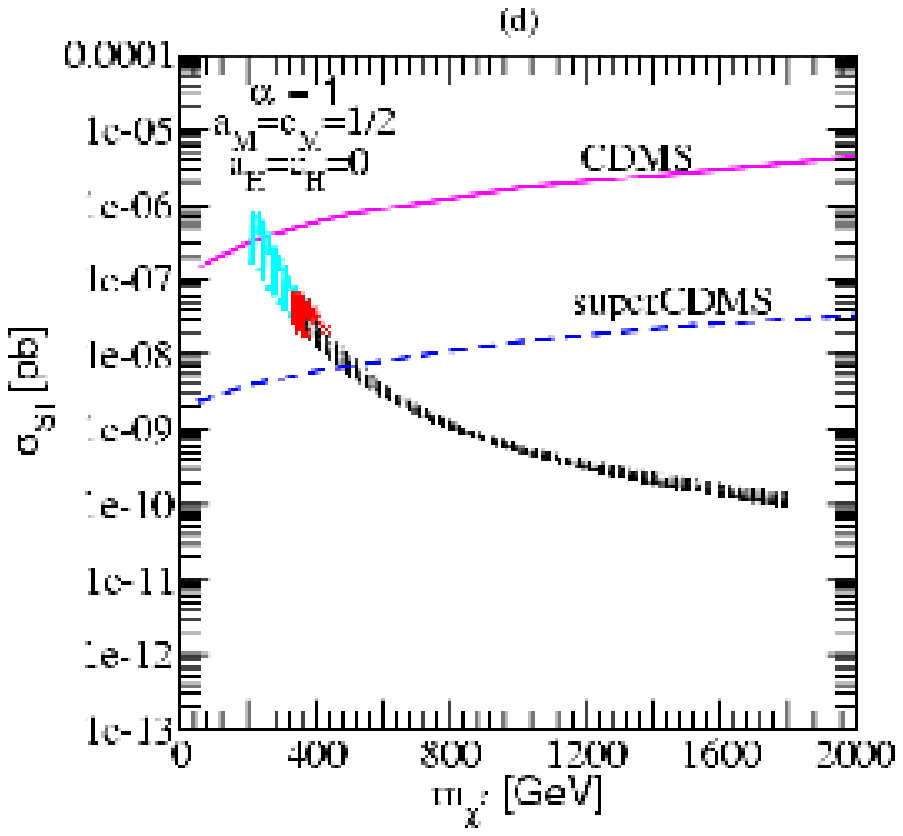}
\end{center}
\vskip -0.4cm \caption{Scatter plot of the spin-independent
neutralino-proton scattering cross section vs. $m_\chi$ in
intermediate scale mirage mediation with $(a_i,c_i)$ specified in
Eq.~(3.2).
}
\label{fig:sigsip}
\end{figure}

If neutralino LSP is the main component of the matter budget in
the Milky Way, it might be detected through the elastic scattering
with terrestrial nuclear target \cite{goodman,lspdm}. In the MSSM,
$t$-channel Higgs boson and $s$-channel squark exchange processes
contribute to the spin-independent (scalar) scattering between
neutralino and nuclei. In many cases, dominant contribution to the
scalar cross section comes from the Higgs exchange process which
becomes  bigger for larger $\tan\beta$, smaller Higgs masses, and
mixed Bino-Higgsino LSP.

In the specific intermediate scale mirage mediation model with
$a_i=c_i=1$, the neutralino LSP is Bino-like and the mass of heavy
CP-even Higgs boson is rather large when we require the neutralino
to be LSP. It is thus expected that the elastic scattering cross
section between neutralino DM and nuclei is rather small. In
Fig.~\ref{fig:sigsip}.a, we depict spin-independent (scalar) cross
section $\sigma_{SI}$ of neutralino-proton scattering as a
function of the LSP neutralino mass in this specific intermediate
scale mirage mediation. Here we have imposed the experimental
bounds on the Higgs/sparticle masses and $b \rightarrow s \gamma$
branching ratio, and required that the lightest neutralino is the
LSP. Red points in the figure correspond to the parameter values
giving the WMAP DM density (\ref{eq:wmap}) under the assumption of
pure thermal production,  while the cyan points represent the
parameter values for which the thermal production mechanism gives
a smaller relic density. As expected, the cross section in the
case of $a_i=c_i=1$ is quite small: $\sigma_{SI}\lesssim 5 \times
10^{-9}$ pb, which is much smaller than the current experimental
upper bound. It is even smaller than the sensitivity of future
experiment such as SuperCDMS \cite{supercdms} which would reach
near $10^{-9}$ pb level. On the other hand, intermediate scale
mirage mediations with different values of $(a_i,c_i)$ have a
quite better prospect for direct detection. As can be seen from
Fig.~\ref{fig:sigsip}, most of the (red) WMAP points are above the
sensitivity of SuperCDMS for the other three cases  of different
$(a_i,c_i)$. This is mainly due to the enhanced Higgsino component
of the neutralino LSP and the reduced Higgs mass.

\begin{figure}[ht!]
\begin{center}
\includegraphics[height=7cm,width=7cm]{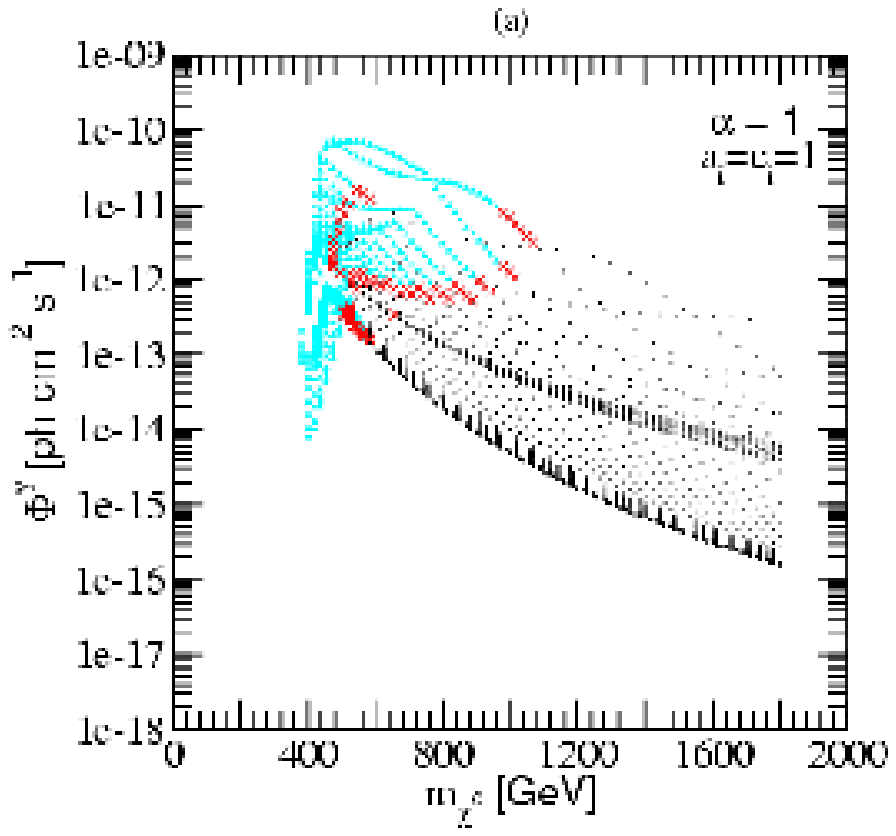}\hskip 1cm
\includegraphics[height=7cm,width=7cm]{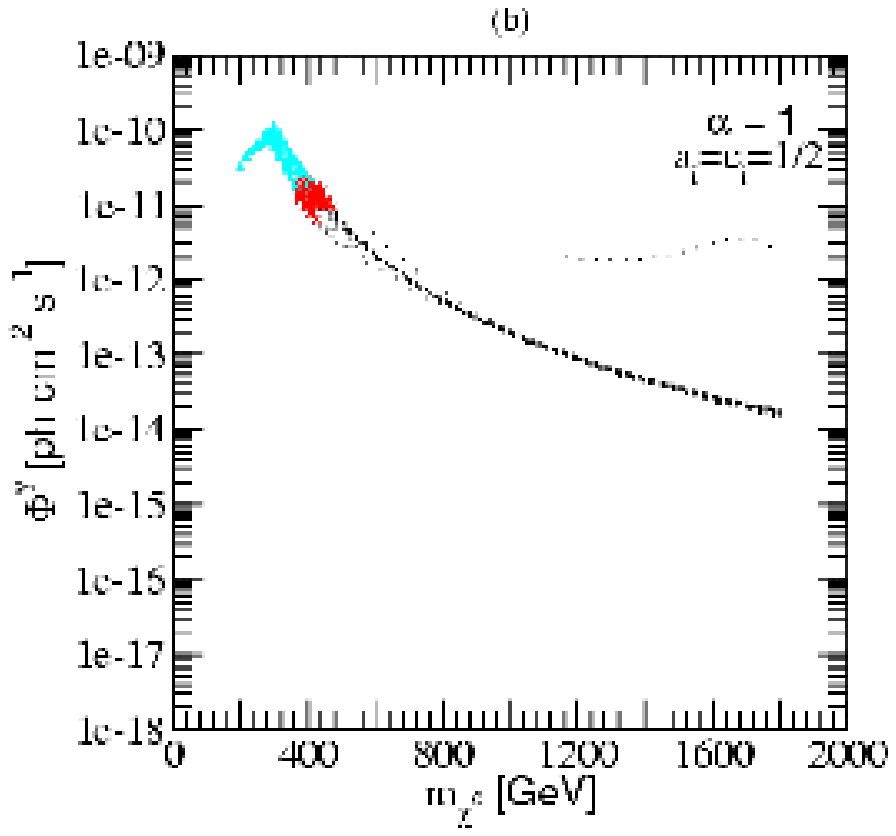}
\vskip 1cm
\includegraphics[height=7cm,width=7cm]{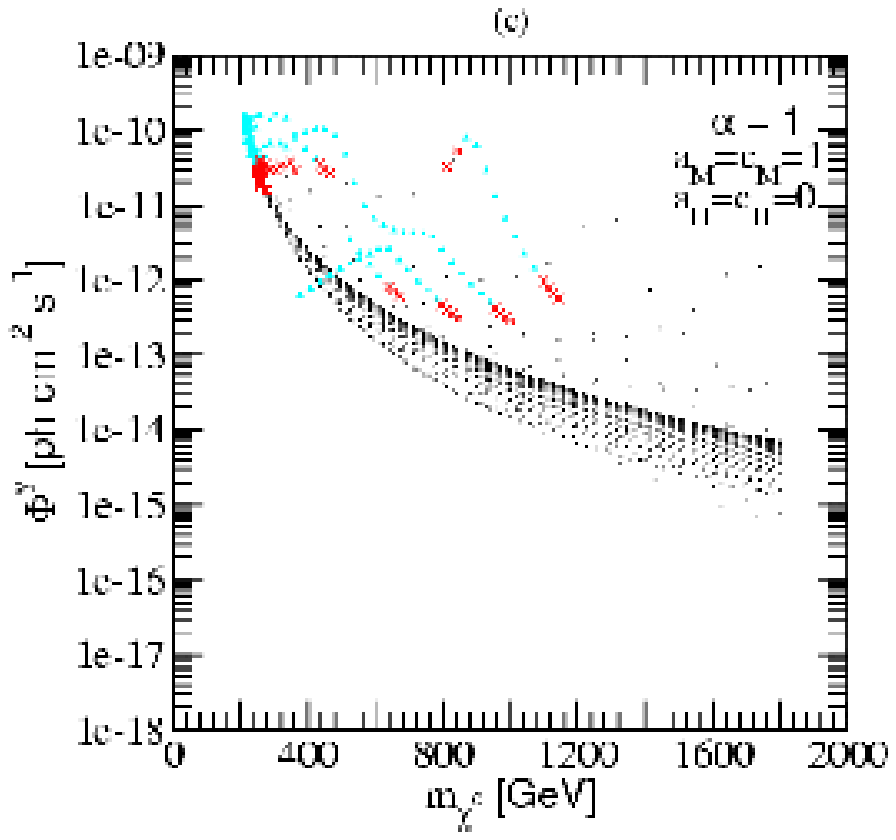}
\hskip 1cm
\includegraphics[height=7cm,width=7cm]{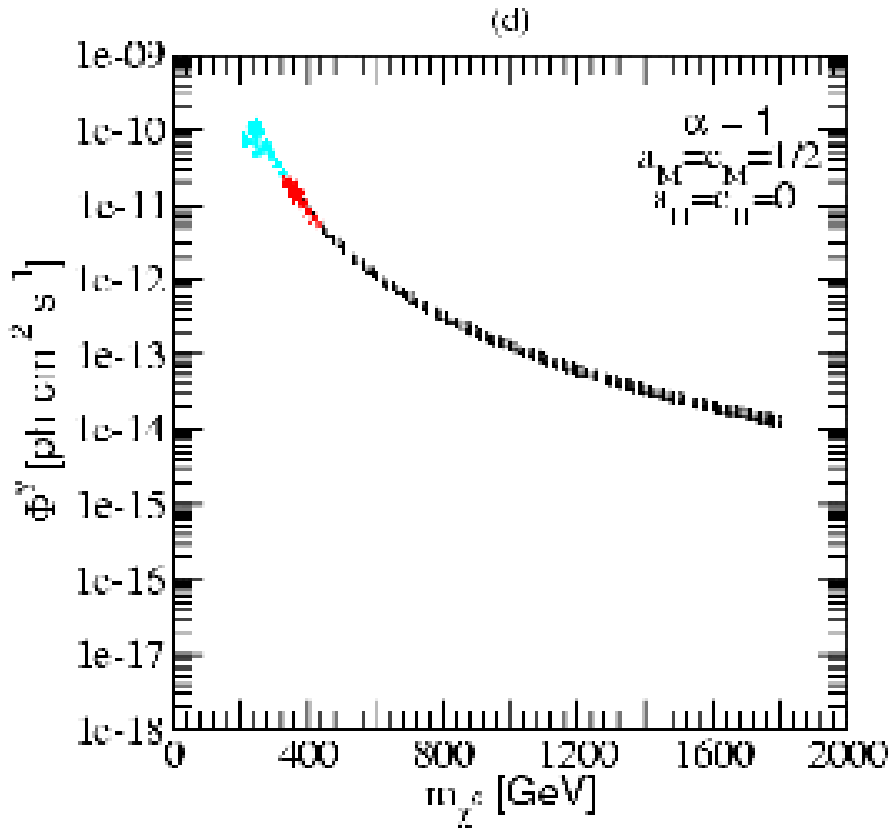}
\end{center}
\vskip -0.5cm \caption{Scatter plot of the continuum gamma ray
flux vs. $m_\chi$ in intermediate scale mirage mediation with
$(a_i,c_i)$ specified in Eq.~(3.2).
}
\label{fig:gamma}
\end{figure}

Let us now examine gamma ray signals  from DM annihilation in the
galactic center,  providing another feasible but indirect
detection method for dark matter. The integrated gamma ray flux
depends on the quantity $\bar{J} (\Delta\Omega)$, which is a
measure of the cuspiness of the galactic halo density profile over
a spherical region of solid angle $\Delta\Omega$. In this paper,
we use a conservative galactic halo model (isothermal halo density
profile) which gives $\bar{J} \sim 30$ with the detector angular
resolution $\Delta\Omega = 10^{-3}$ sr and set $E_{thr} = 1$ GeV
for gamma ray energy threshold. Fig.~\ref{fig:gamma} shows
continuum gamma ray flux from the galactic center in intermediate
scale mirage mediation scenarios under consideration, where red
points give the WMAP value (\ref{eq:wmap}) of the relic DM
density. Here the four different choices of $a_i=c_i$ do not lead
to a dramatic difference in the gamma ray flux. The maximal value
of flux given by the most favored WMAP (red) points is about ${\rm
few}\times 10^{-11} \rm{cm^{-2} s^{-1}}$ which is somewhat below
the expected reach ($\sim 10^{-10}\rm{cm^{-2} s^{-1}}$) of GLAST,
although the (cyan) points giving smaller relic density can give a
larger flux around $10^{-10}\rm{cm^{-2} s^{-1}}$. However, it
should be noticed that our calculation for the gamma ray flux is
based on a conservative halo density profile. If one uses an
extreme halo model like the spiked profile \cite{moore}, the
resulting gamma ray flux increases by a factor of $\sim 10^4$. In
this case, the gamma ray signals can be detected for a significant
portion of the parameter space. A caveat is that the continuum
gamma ray signals suffer from unknown astrophysical background.
Recent observations of a bright gamma ray source in the direction
of galaxy center by the Air Cherenkov Telescopes such as H.E.S.S.
\cite{hess} might be explained by an astrophysical process rather
than the dark matter annihilation \cite{hooper}.

\begin{figure}[ht!]
\begin{center}
\includegraphics[height=7cm,width=7cm]{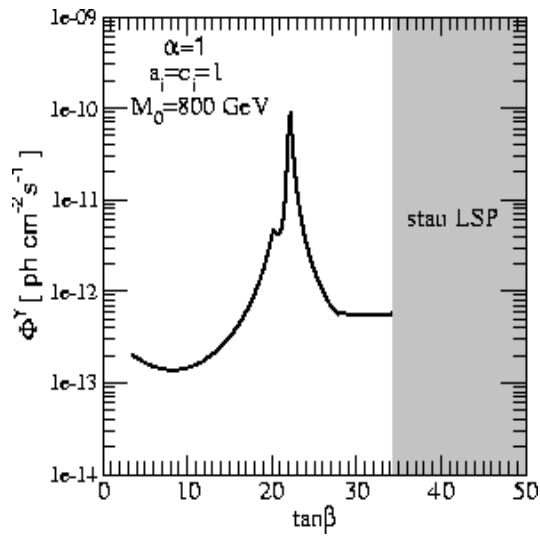}
\end{center}
\vskip -0.5cm \caption{Continuum gamma ray flux as a function of
$\tan\beta$ in the intermediate scale mirage mediation with
$a_i=c_i=1$ and $M_0=800$ GeV. Note the resonant peak due to the
psuedo-scalar Higgs resonance.} \label{fig:gammas}
\end{figure}

We finally notice an interesting enhancement of the gamma ray flux
due to the Higgs resonance effect. Fig.~\ref{fig:gammas} shows the
gamma ray flux from the galactic center as a function of $\rm
tan\beta$ in the specific intermediate scale mirage mediation with
$a_i=c_i=1$ and $M_0=800$ GeV. One can see a clear enhancement of
the flux around $\rm tan\beta \sim 22$ for which $m_A \sim 2
m_\chi$. In this case, neutralino annihilation to heavy quarks is
dominated and the subsequent quark hadronization produces many
gamma rays.

%


\section{Neutralino DM for generic mirage messenger scale}


In the previous section, we have examined the prospect of
neutralino DM in intermediate scale mirage mediation models
($\alpha=1$). As was discussed in section 2, in string
compactifications with non-trivial dilaton-modulus mixing, the
anomaly to modulus mediation ratio $\alpha$ can have a more
variety of values. In fact, the nature of neutralino LSP is
somewhat sensitive to the value of $\alpha$, typically it changes
from Bino-like to Higgsino-like via Bino-Higgsino mixing region
when $\alpha$ is increased from zero to a value of order unity.
This feature is essentially due to the following behavior of the
gaugino masses  as a function of $\alpha$: \bea M_3:M_2:M_1\simeq
 (1-0.3\alpha)g_3^2:(1+0.1\alpha)g_2^2:(1+0.66\alpha)g_1^2,
 \eea
If $\alpha$ increases from zero, the gluino mass decreases as
$M_3\propto (1-0.3\alpha)$. Smaller $M_3$ then weakens  the
radiative electroweak symmetry breaking mechanism as it gives a
smaller stop mass-square, thus leads to smaller $|m_{H_u}|^2$ and
$|\mu|$ at the weak scale. On the other hand,
 the Bino mass increases as $M_1\propto
(1+0.66\alpha)$, thus the lightest neutralino changes from
Bino-like to Higgsino-like when $\alpha$ is varying from zero to a
positive value of order unity. If $\alpha$ is further increased,
eventually the model does not allow electroweak symmetry breaking.
In this section, we extend the analysis of the previous section to
the range of $\alpha$ from zero to the value at which the
electroweak symmetry starts to be restored.

\subsection{Parameter region of neutralino LSP and thermal relic
Density}

%

\begin{figure}[ht!]
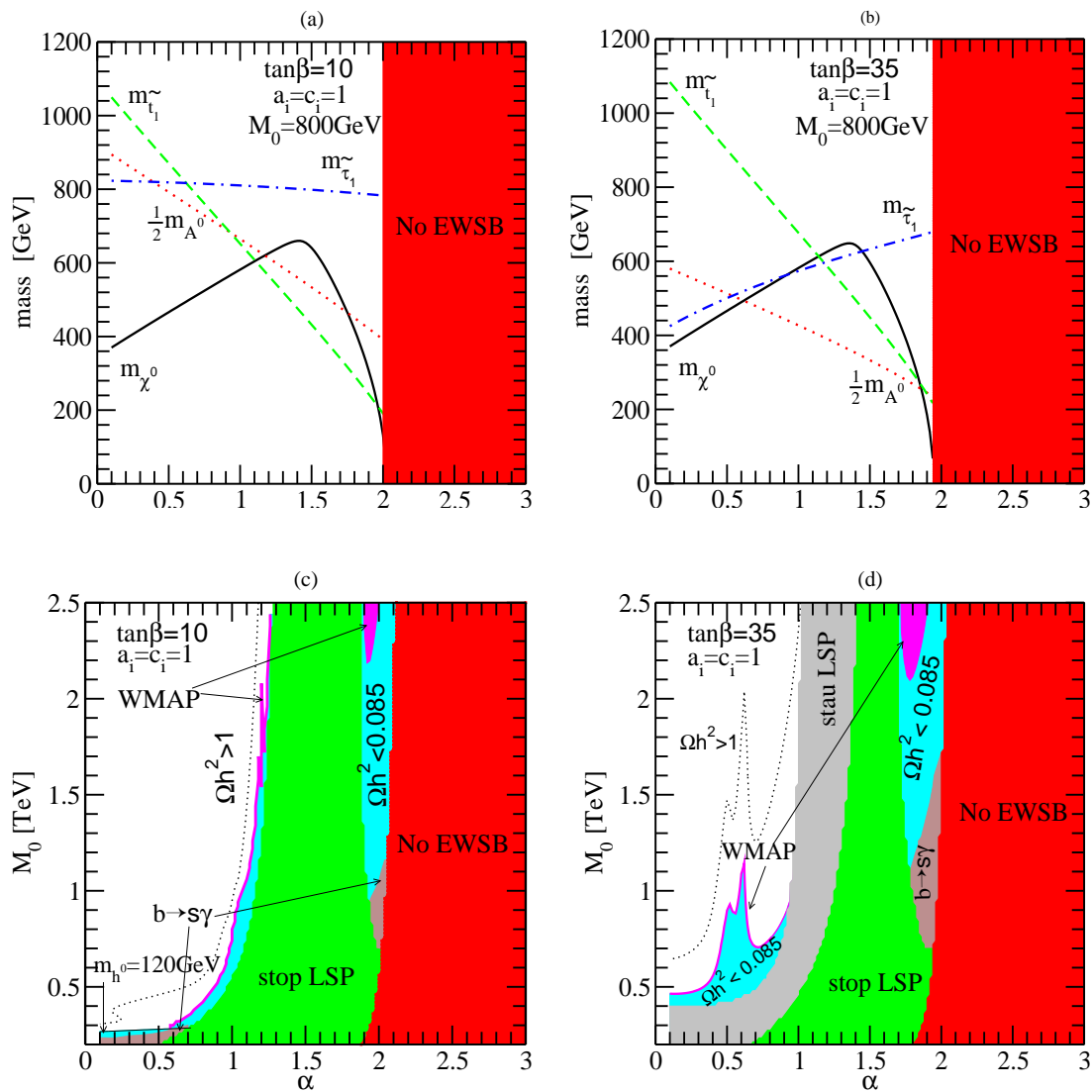

\begin{center}
\includegraphics[height=7cm,width=7cm]{tb10-mass.eps}\hskip 0.5cm
\includegraphics[height=7cm,width=7cm]{tb35-mass.eps}
\vskip 0.5cm
\includegraphics[height=7cm,width=7cm]{tb10.eps}\hskip 0.5cm
\includegraphics[height=7cm,width=7cm]{tb35.eps}
\end{center}
\vskip -0.4cm \caption{Sparticle masses vs. $\alpha$ for
$\tan\beta=10$ and $\rm tan\beta=35$ in case with $a_i=c_i=1$. The
lower figures show the parameter space of neutralino LSP and its
thermal relic density on the plane of ($\alpha$, $M_0$).}
\label{fig:density-tb35}
\end{figure}

Again, let us first consider the case with $a_i=c_i=1$. We will
treat $M_0$ and $\alpha$ as free parameters, while focusing on
$\tan\beta=10$ and 35. Figs.~\ref{fig:density-tb35}.a and
\ref{fig:density-tb35}.b  show how some of the superparticle
masses vary as a function of $\alpha$ for a fixed $M_0 = 800$ GeV.
For $\alpha\lesssim 1$, the LSP is the lightest neutralino which
is mostly Bino, and thus its mass varies as $m_{\chi^0}\simeq
M_1\propto (1+0.66\alpha)$. In the range of
$1\lesssim\alpha\lesssim 1.8$, stau or stop becomes the LSP. For
$1.8\lesssim \alpha\lesssim 2$, the lightest neutralino which is
now mostly Higgsino becomes  the LSP. If $\alpha$ increases
further, the model does not allow  electroweak symmetry breaking.

In Figs.~\ref{fig:density-tb35}.c and \ref{fig:density-tb35}.d,
the two distinct magenta regions seperated by stop/stau LSP
regions give the WMAP DM density, $0.085<\Omega_{DM} h^2<0.119$,
under the assumption that all neutralino DMs are produced by the
conventional thermal production mechanism. Below (above) these
magenta regions, $\Omega_\chi h^2 < 0.085$  ($> 0.119$). In the
Bino-like LSP region, stop-neutralino coannihialtion plays a
crucial role to get the WMAP DM density for $\rm tan\beta=10$,
while stau-neutralino coannihilation or pseudoscalar Higgs
resonance processes are important for $\rm tan\beta=35$. For
Higgsino-like LSP, the charged Higgsino $\chi_1^\pm$ and two
neutral Higgsinos $\chi_1^0, \chi_2^0$ are nearly degenerate. Then
the dominant annihilation processes are the neutralino pair
annihilation into gauge bosons, and the neutralino-chargino
co-annihilation into fermion pair \cite{coanil}. These
annihilations of Higgsino-like LSP are very efficient, so that the
relic mass density is too small unless $m_\chi^0$ is quite heavy.
Indeed, from Figs.~\ref{fig:density-tb35}.c and
\ref{fig:density-tb35}.d, we can see that the WMAP DM density is
obtained only for $M_0\gtrsim 2.2$ TeV in the Higgsino LSP region
around $\alpha\sim 1.8$. However it should be stressed that the
cyan regions of Figs.~\ref{fig:density-tb35}.c and
\ref{fig:density-tb35}.d can be allowed  if some part of DM were
produced by non-thermal mechanism. Such parameter region contains
$\alpha\sim 1.8$ and $M_0\sim 1$ TeV for which the neutral
Higgsino with $m_{\chi^0}\sim 200$ GeV is the LSP and the stop is
rather light as $m_{\tilde{t}_1}\sim 250$ GeV.  
The brown regions are excluded by the $b \to s \gamma$ constraint. On the brown
region in Fig. 9.c, the chargino loop
contribution to $b\to s\gamma$ dominates, which results in Br$(b
\to s \gamma)$ smaller than the experimentally allowed range. 

\vskip 0.8cm

\begin{figure}[ht!]
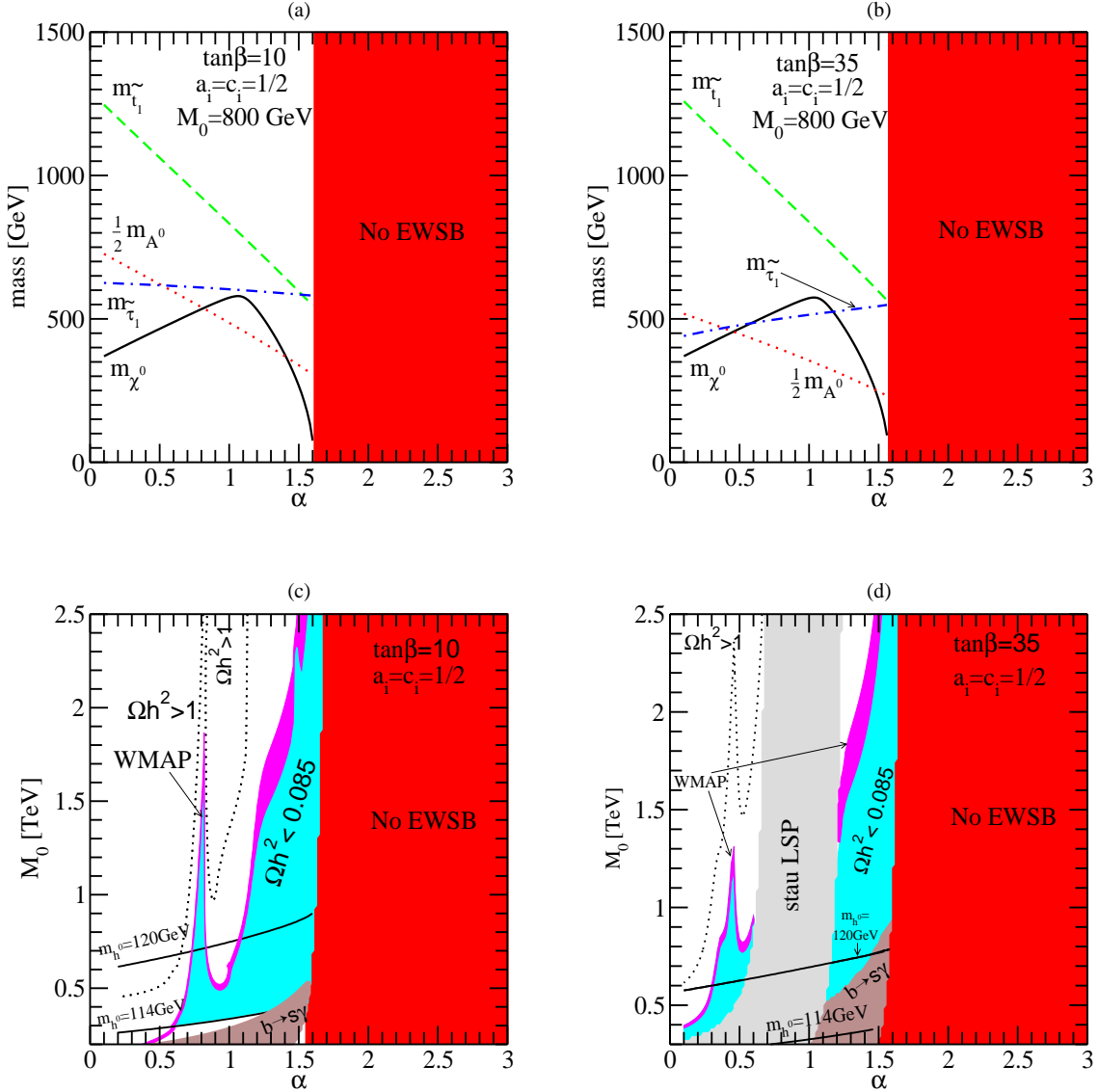

\begin{center}
\includegraphics[height=7cm,width=7cm]{nzmw-tb10-mass.eps}\hskip 1cm
\includegraphics[height=7cm,width=7cm]{nzmw-tb35-mass.eps}
\vskip 1cm
\includegraphics[height=7cm,width=7cm]{nzmw-tb10.eps}\hskip 1cm
\includegraphics[height=7cm,width=7cm]{nzmw-tb35.eps}
\end{center}
\vskip -0.4cm
\caption{The results for the case in which $a_i=c_i=1/2$ for both
the Higgs and matter multiplets.} \label{fig:density-nzero}
\end{figure}
\begin{figure}[ht!]
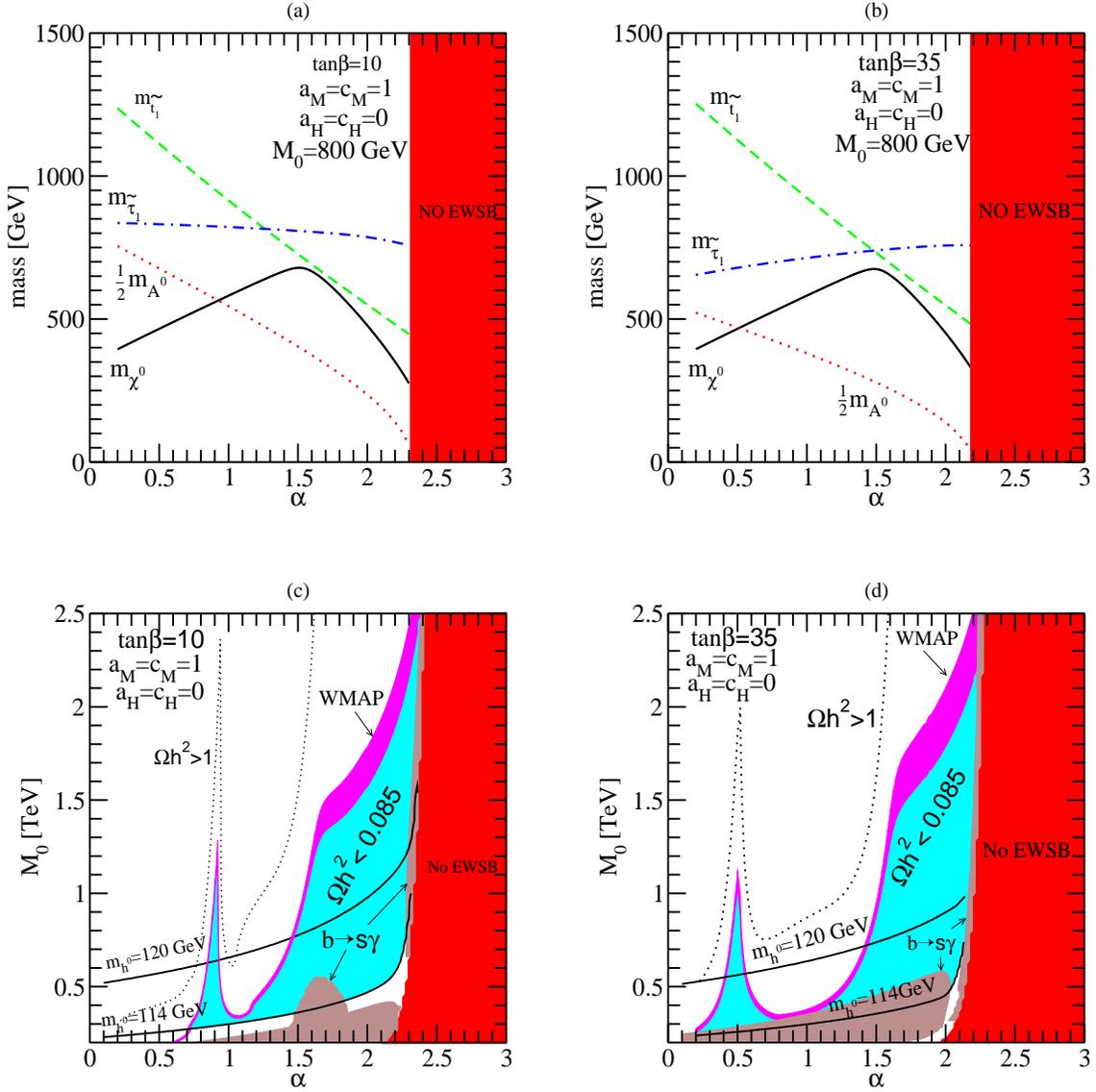

\begin{center}
\includegraphics[height=7cm,width=7cm]{nzmw2-tb10-mass.eps}\hskip 1cm
\includegraphics[height=7cm,width=7cm]{nzmw2-tb35-mass.eps}\vskip
1cm
\includegraphics[height=7cm,width=7cm]{nzmw2-tb10.eps}\hskip 1cm
\includegraphics[height=7cm,width=7cm]{nzmw2-tb35.eps}
\end{center}
\vskip -0.4cm \caption{The results for the case in which $a_{\rm
H}=c_{\rm H}=0$ and $a_{\rm M}=c_{\rm M}=1$.}
\label{fig:density-nzero2}
\end{figure}

Let us now consider the case with  $a_i=c_i=1/2$. Obviously,
smaller $(a_i,c_i)$ give smaller stop/stau mass-squares at
$M_{GUT}$. However, as was anticipated in the previous section,
$X_I$ ($I=t,b,\tau$) which govern the RG evolution of stop/stau
mass-squares (see Eq.~\ref{rgequation}) become smaller also, which
would increase the stop/stau masses at the weak scale. Together
with the reduction of $A_{H_uQ_3U_3}$, this effect on the RG
evolution eventually makes the physical lighter stop mass
$m_{\tilde t_1}$ larger compared to the case with $a_i=c_i=1$. On
the other hand, stau masses are more affected by the change of the
boundary values, thus their weak scale values become lighter
compared to the case of $a_i=c_i=1$.
 Smaller $X_t$ leads to
also a smaller $|m_{H_u}^2|$ at the weak scale, resulting the
reduction of the Higgsino mass $\mu$ and the pseudoscalar Higgs
boson mass $m_A$. Figs.~\ref{fig:density-nzero}.a and
\ref{fig:density-nzero}.b show all of these features.  Again, as
$\alpha$ increases, the neutralino LSP changes from Bino-like to
Higgsino-like. Comparing to Fig.~\ref{fig:density-tb35}, the
lighter stop becomes heavier, while the lighter stau and the
pseudoscalar Higgs become lighter. As a consequence, the stop LSP
region disappears, but the stau LSP region at large $\tan\beta$
becomes larger. The magenta regions of
Figs.~\ref{fig:density-nzero}.c and \ref{fig:density-nzero}.d
correspond to the parameter region giving the WMAP DM density
under the assumption of pure thermal production. They clearly show
the Higgsino-like LSP at $\alpha>1$ and also the pseudoscalar
Higgs resonance effect for the Bino-like LSP at smaller $\alpha$.

\vskip 0.8cm
\begin{figure}[ht!]
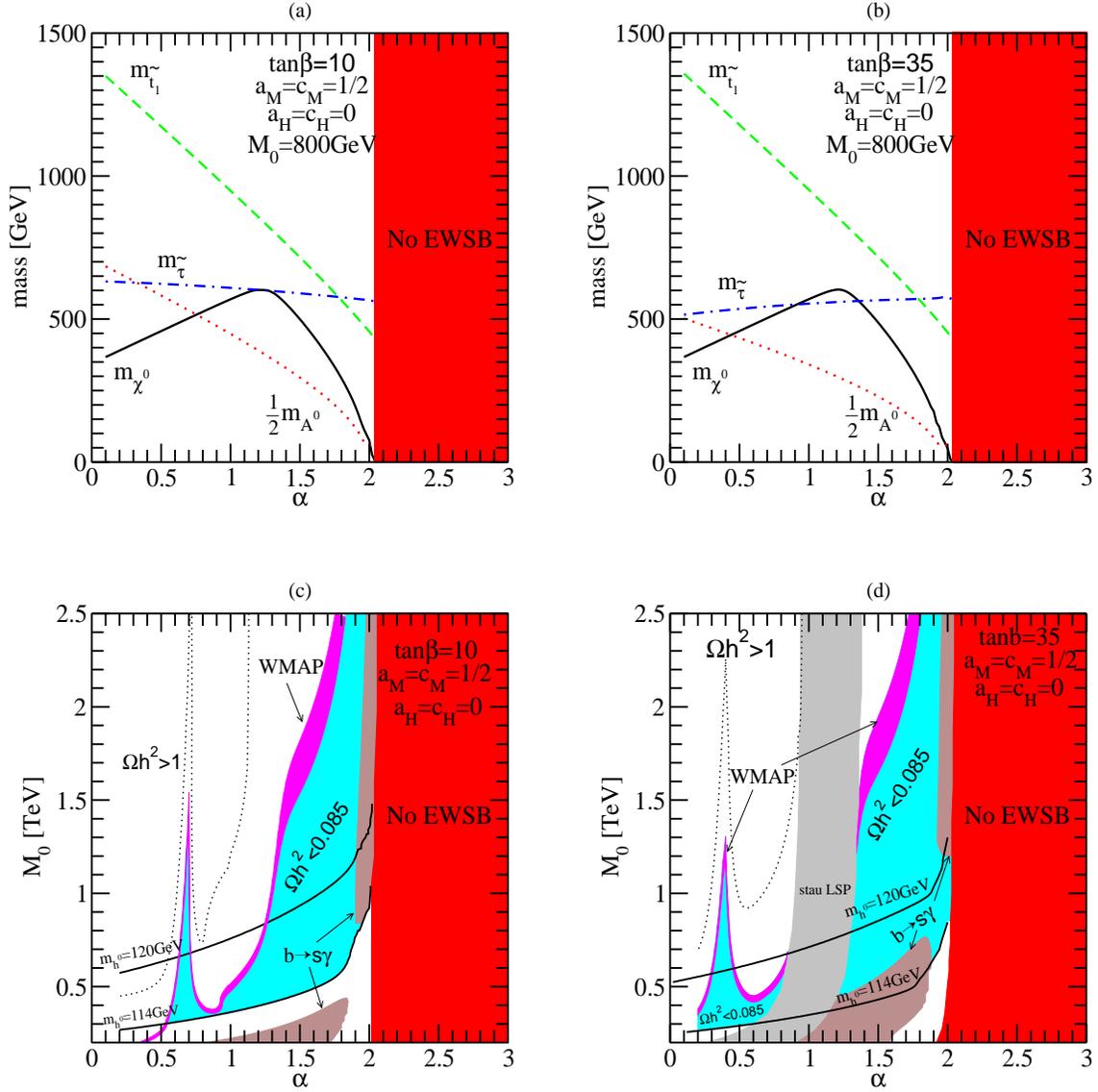

\begin{center}
\includegraphics[height=7cm,width=7cm]{nzmw3-tb10-mass.eps}\hskip 1cm
\includegraphics[height=7cm,width=7cm]{nzmw3-tb35-mass.eps} \vskip
1cm
\includegraphics[height=7cm,width=7cm]{tb10-nq05-nh1.eps}\hskip 1cm
\includegraphics[height=7cm,width=7cm]{tb35-nq05-nh1.eps}
\end{center}
\vskip -0.3cm \caption{The results for the case in which $a_{\rm
H}=c_{\rm H}=0$ and $a_{\rm M}=c_{\rm M}=1/2$.}
\label{fig:density-nq05-nh1}
\end{figure}

Fig.~\ref{fig:density-nzero2} shows the results for the case in
which $a_{\rm H}=c_{\rm H}=0$ for the Higgs multiplets, while
$a_{\rm M}=c_{\rm M}=1$ for the quark/lepton matter multiplets. A
characteristic feature of this case is that the lightest
neutralino is the LSP over the entire region of parameter space
allowing the electroweak symmetry breaking. Compared to the case
in which  $a_i=c_i=1$ for both the Higgs and matter multiplets,
$X_I$ ($I=t,b,\tau$) for the RG evolution (\ref{rgequation})  have
smaller values, while the boundary values of stop/stau
mass-squares remain the same. This results in heavier stop and
stau at the weak scale. Except for the absence of stop/stau LSP
region, other features are somewhat similar to other cases.
The brown regions are excluded by the $b \to s \gamma$ constraint. 
On the brown region with small $M_0$ in Fig. 11.c, the chargino loop
contribution to $b\to s\gamma$ dominates, which results in Br$(b
\to s \gamma)$ smaller than the experimentally allowed range. 
On the other hand, the charged Higgs boson loop becomes significant
in the large $M_0$ region, making  the predicted Br$(b \to s
\gamma)$ exceed the experimental bound. The region between those
two brown regions is allowed due to the cancellation between the
chargino and charged Higgs boson loop contributions.

Finally, Fig.~\ref{fig:density-nq05-nh1} is for the case with
$a_H=c_H=0$ and $a_{\rm M}=c_{\rm M}=1/2$. The results are quite
similar to the case in which $a_i=c_i=1/2$ for both the Higgs and
matter multiplets. The $\alpha=2$ region of this case corresponds
to the TeV scale mirage mediation model proposed in
\cite{tevmirage} as a model to minimize the fine tuning for the
electroweak symmetry breaking in the MSSM.

\subsection{Dark matter detections} \vspace{0.5cm}

\begin{figure}[ht!]
\begin{center}
\includegraphics[height=6cm,width=6cm]{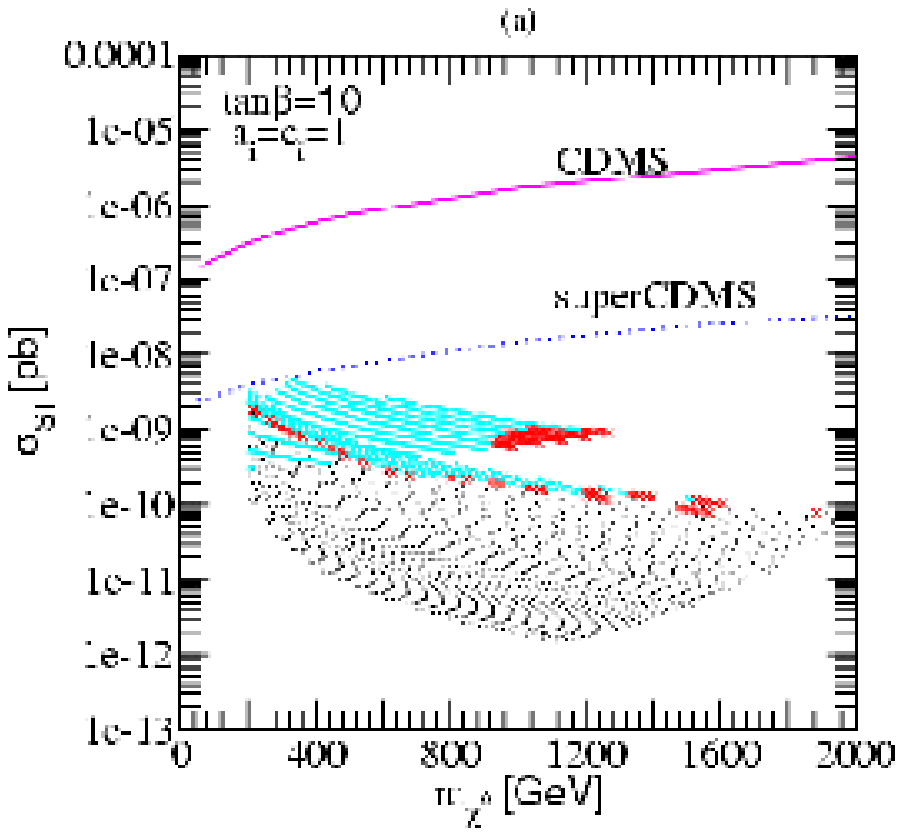}\hskip 1.0cm
\includegraphics[height=6cm,width=6cm]{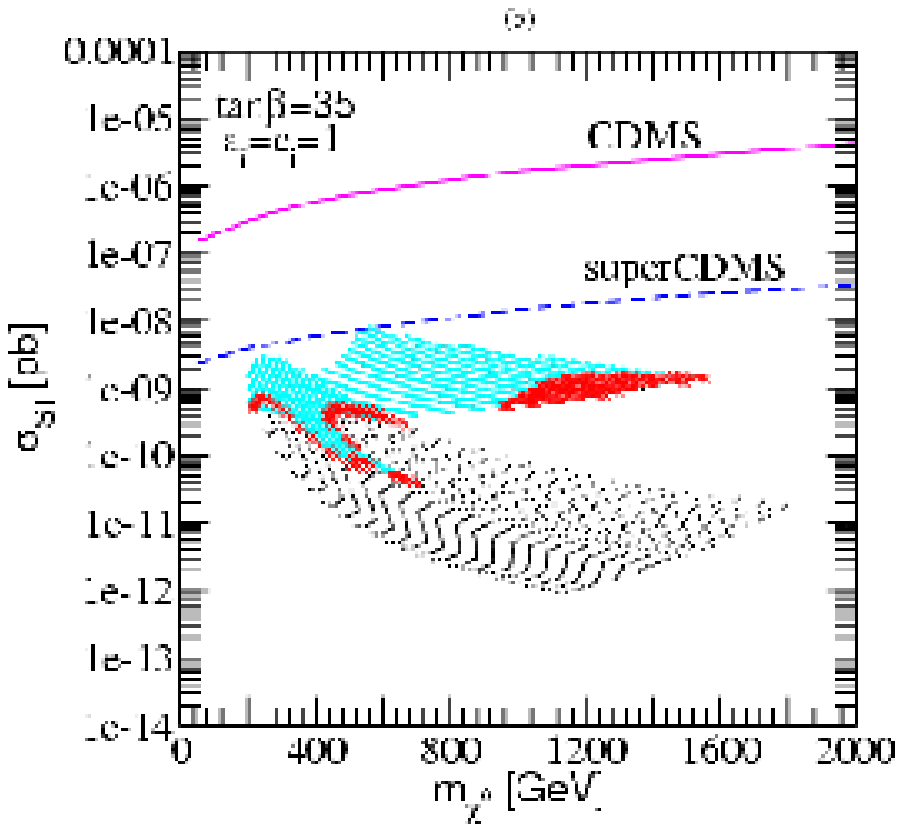}\vskip 0.5cm
\end{center}
\vskip -0.5cm \caption{Spin-independent neutralino and proton
scattering cross section in case with $a_i=c_i=1$.}
\label{fig:sigsip-tb35}
\end{figure}

To see the prospect of direct DM detection, spin-independent cross
section  of the neutralino-proton scattering is presented in
Fig.~\ref{fig:sigsip-tb35} for the case with $a_i=c_i=1$. Here, we
imposed the experimental bounds on sparticle and Higgs masses, and
$b\rightarrow s\gamma$ branching ratio. In the figures, the red
points give the WMAP DM density: $0.085 < \Omega_\chi h^2 <
0.119$, the cyan  corresponds to the region giving  $\Omega_\chi
h^2< 0.085$, and the rest gives $\Omega_\chi h^2 > 0.119$, under
the assumption of pure thermal production of neutralino DM.
One can notice that there are two distinct branches  of the WMAP
points which correspond to the Bino branch and the Higgsino branch, respectively.
In our scan, Higgsino-like LSP gives a larger $\sigma_{SI}$ for a given
$m_\chi$. The dominant contribution to $\sigma_{SI}$ usually comes
from the Higgs exchange process which becomes
 significant if the LSP neutralino is a mixed Bino-Higgsino state.
On the other hand, for $a_i=c_i=1$, the LSP neutralino is either
Bino-like or Higgsino-like since the mixed Bino-Higgsino region
gives a stop or stau LSP.  Therefore, it is expected that the
cross section is rather small for the case with $a_i=c_i=1$.
Indeed, Fig.~\ref{fig:sigsip-tb35} shows that the predicted values
are all less than the current and near future experimental
sensitivity.

\begin{figure}[ht!]
\begin{center}
\includegraphics[height=6cm,width=6cm]{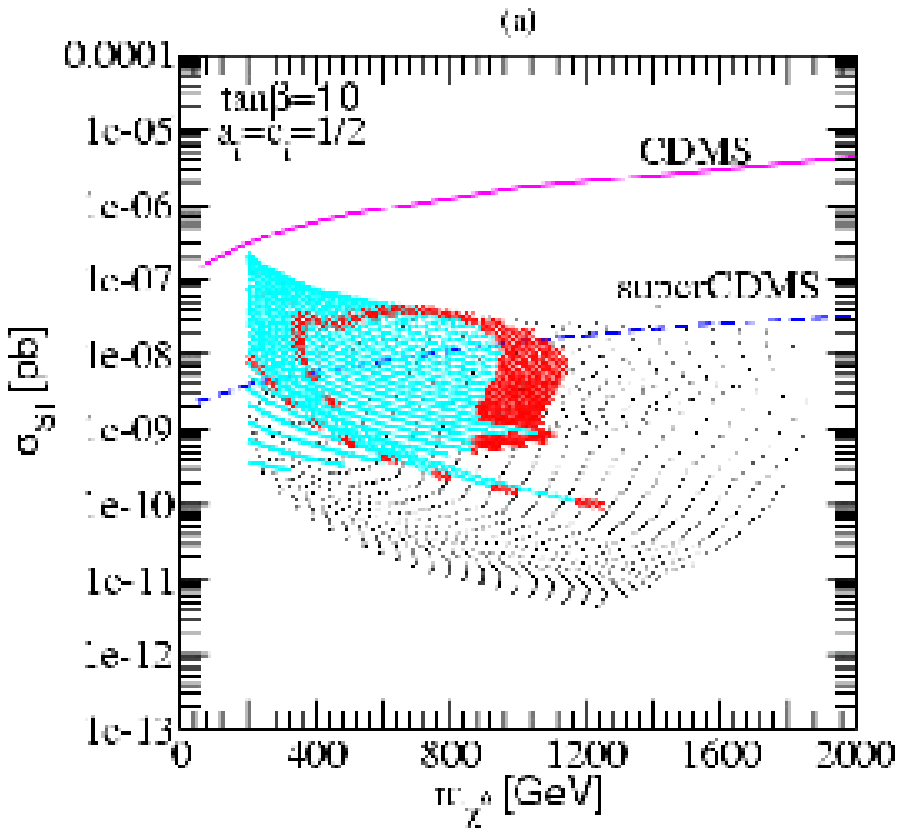}\hskip 1cm
\includegraphics[height=6cm,width=6cm]{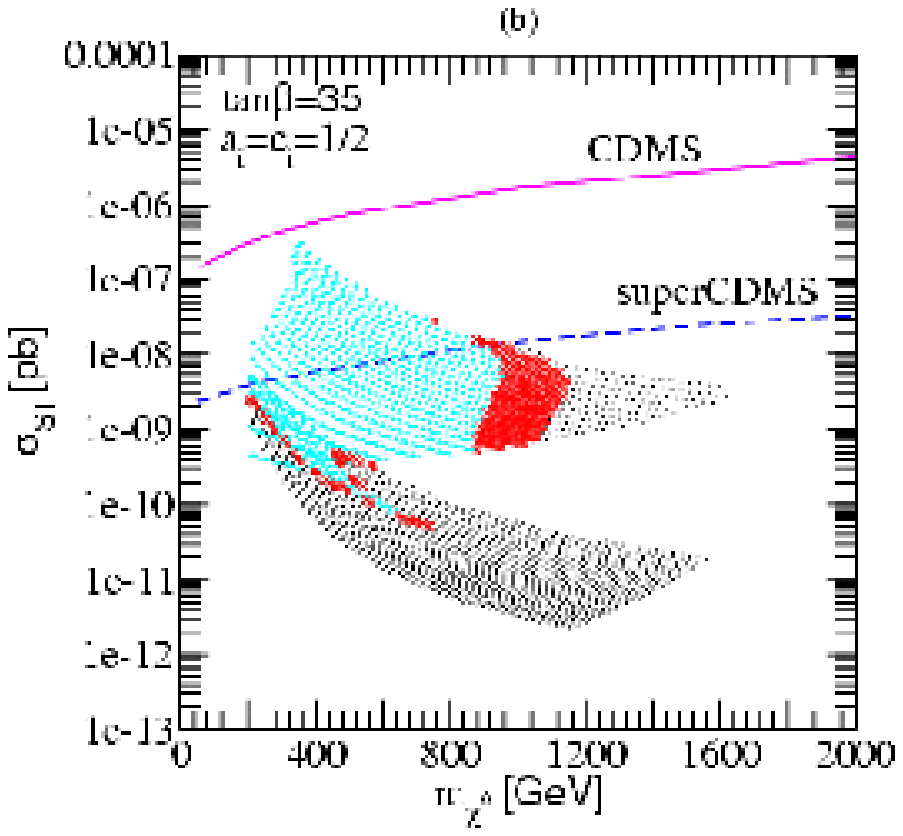}\vskip
0.5cm
\includegraphics[height=6cm,width=6cm]{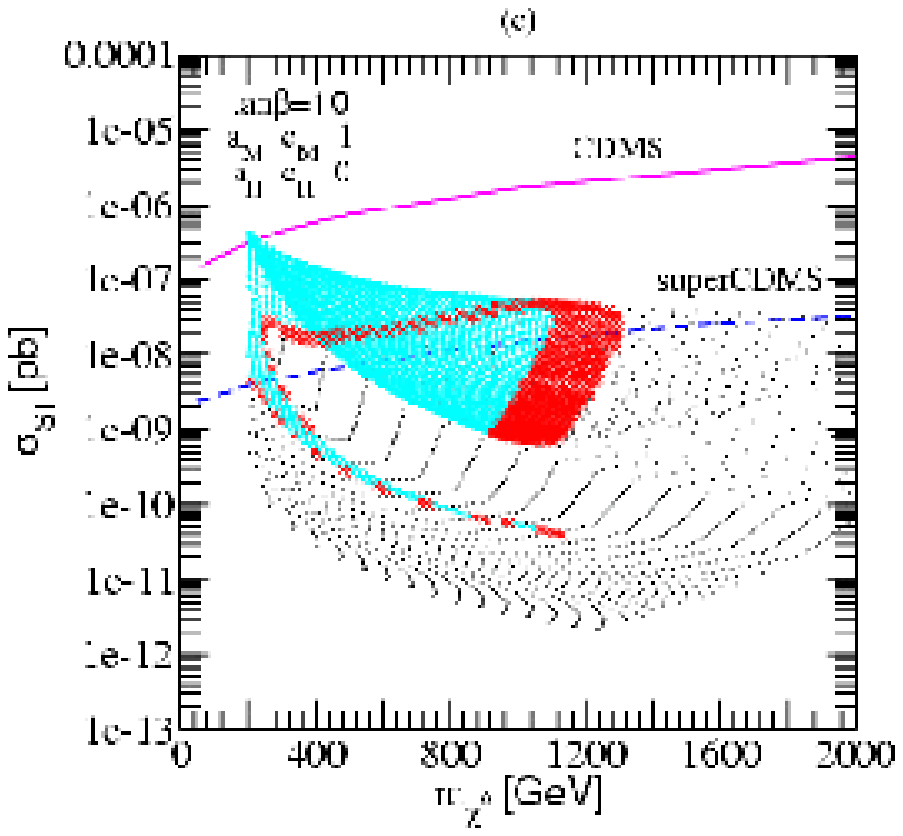}\hskip 1cm
\includegraphics[height=6cm,width=6cm]{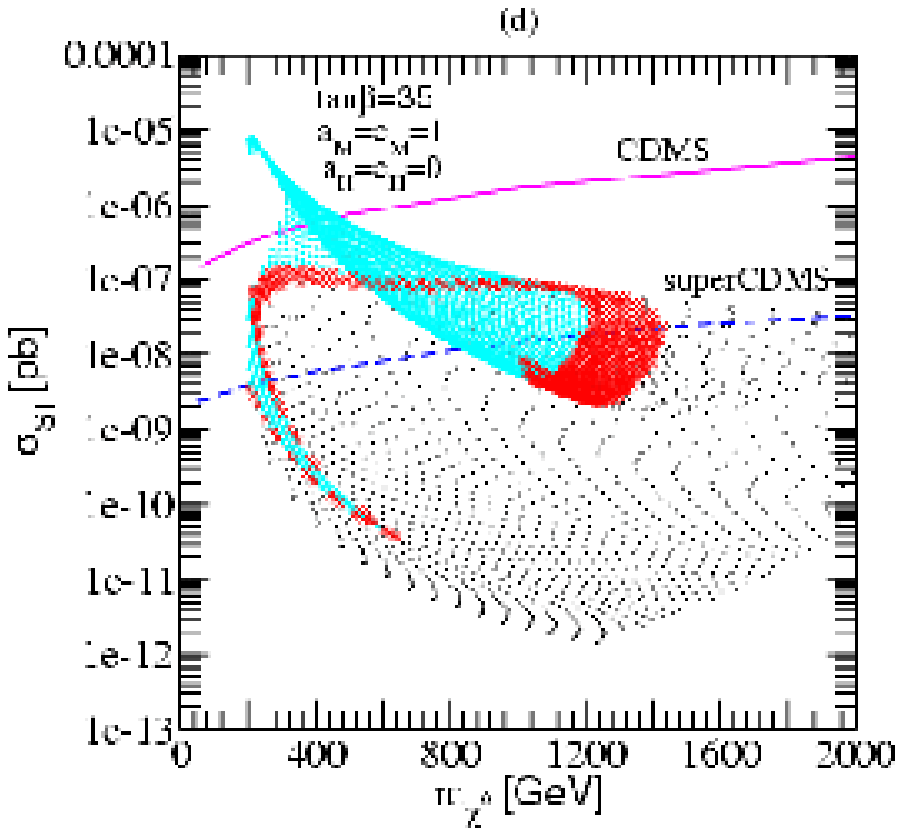}\vskip 0.5cm
\includegraphics[height=6cm,width=6cm]{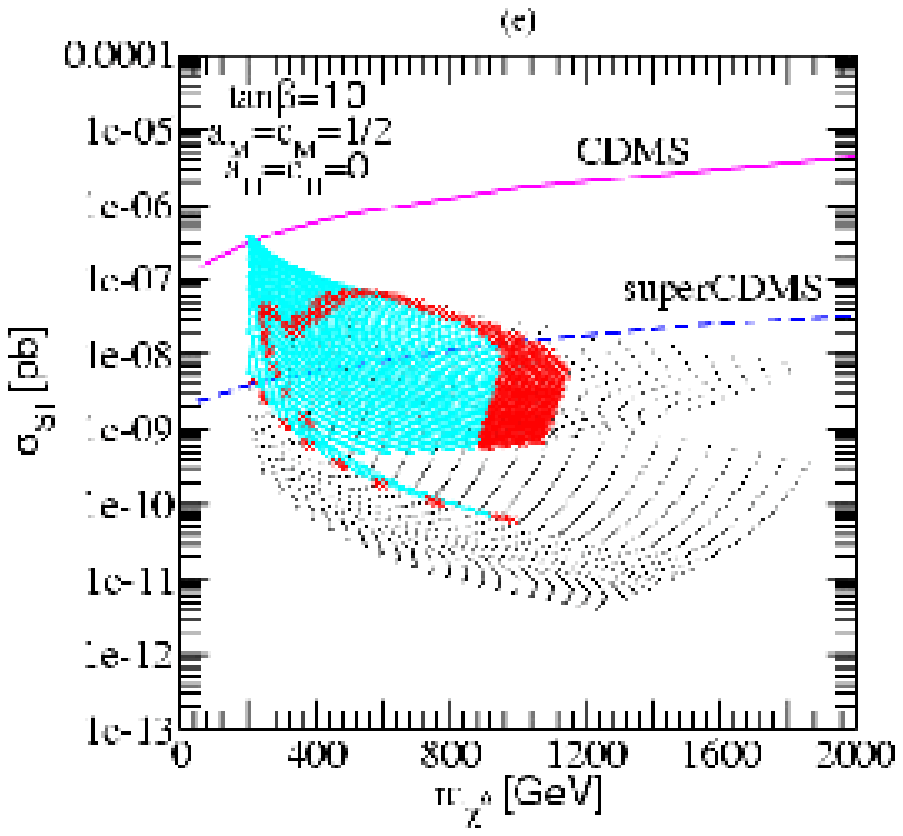}\hskip 1cm
\includegraphics[height=6cm,width=6cm]{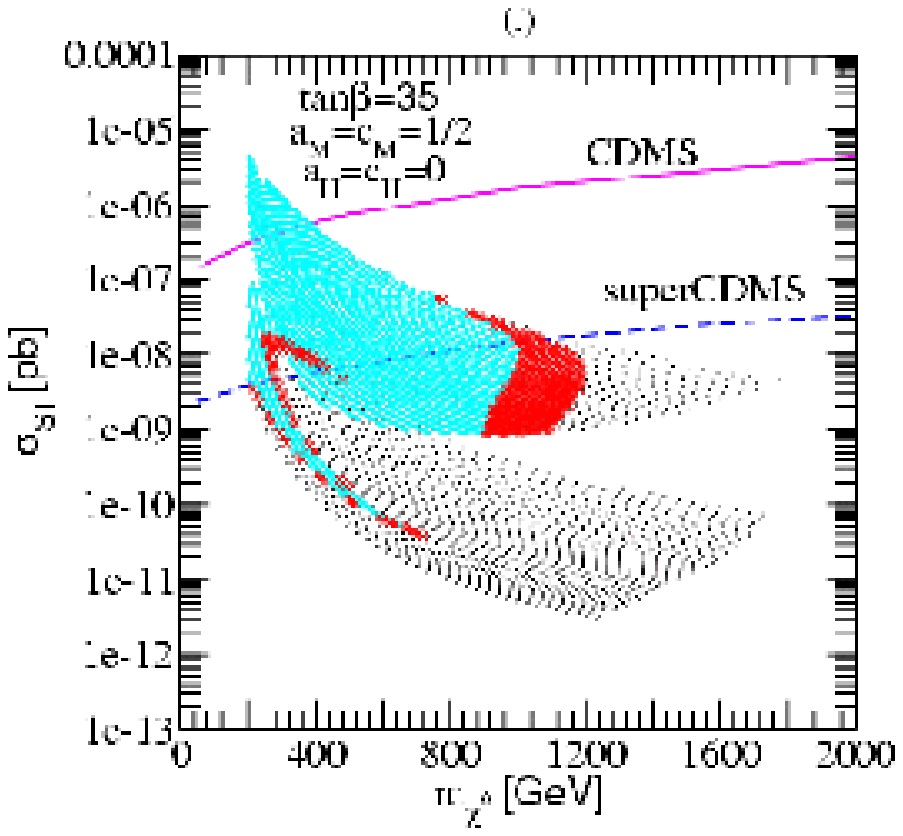}\vskip 0.5cm
\end{center}
\vskip -0.5cm \caption{Spin-independent neutralino and proton
scattering cross section for other values of $(a_i,c_i)$ giving a
mixed Bino-Higgsino LSP over a significant fraction of the
parameter space.} \label{fig:sigsip-tb36}
\end{figure}

However, the prospect of direct DM detection is dramatically
changed if one considers other choices of $a_i$ and $c_i$.
Fig.~\ref{fig:sigsip-tb36} shows the predictions for
spin-independent cross section of the neutralino-proton scattering
for the three other choices of $(a_i,c_i)$  giving a mixed
Bino-Higgsino LSP over a significant fraction of the parameter
space and also a reduced value of the pseudoscalar Higgs mass.
These values of $a_i$ and $c_i$ give  heavier stop, thereby the
$b\rightarrow s\gamma$ constraint becomes less significant
compared to the case with $a_i=c_i=1$. Again, the red points
represent the parameter values giving the WMAP DM density $ 0.085
< \Omega_\chi h^2 < 0.119$, the cyan points give $\Omega_\chi
h^2 < 0.085$, and the rest stands for $\Omega_\chi h^2 > 0.119$,
under the assumption of thermal production of neutralino LSP.
As expected, the scattering cross sections are largely enhanced
compared to the case with $a_i=c_i=1$. Now, much of the WMAP
points give $\sigma_{SI}$ exceeding the sensitivity limit of the
planned SuperCDMS experiment. If one includes the cyan points,
the cross section can be much bigger, reaching even at the current
CDMS sensitivity limit.

\begin{figure}[ht!]
\begin{center}
\includegraphics[height=6cm,width=6cm]{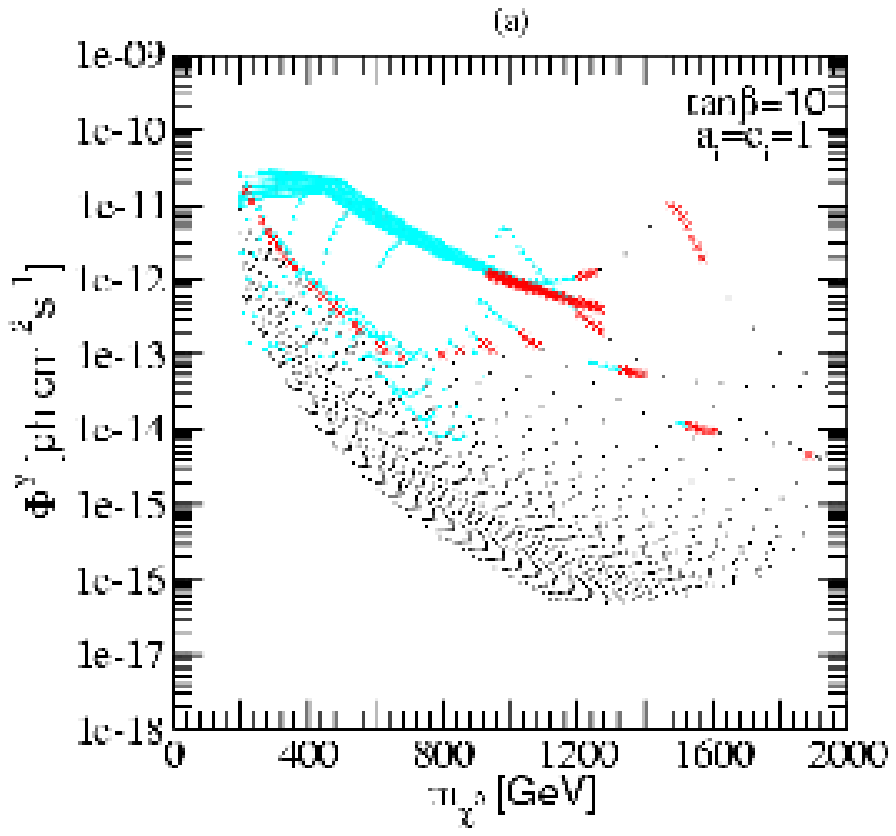}\hskip 0.5cm
\includegraphics[height=6cm,width=6cm]{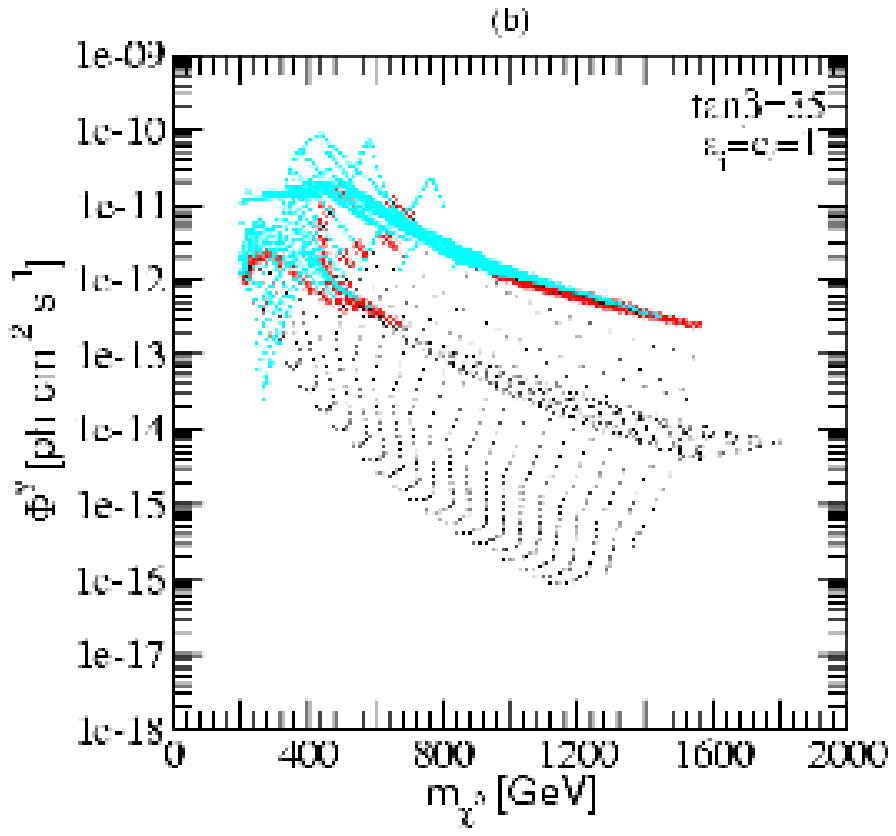}\vskip 1.0cm
\includegraphics[height=6cm,width=6cm]{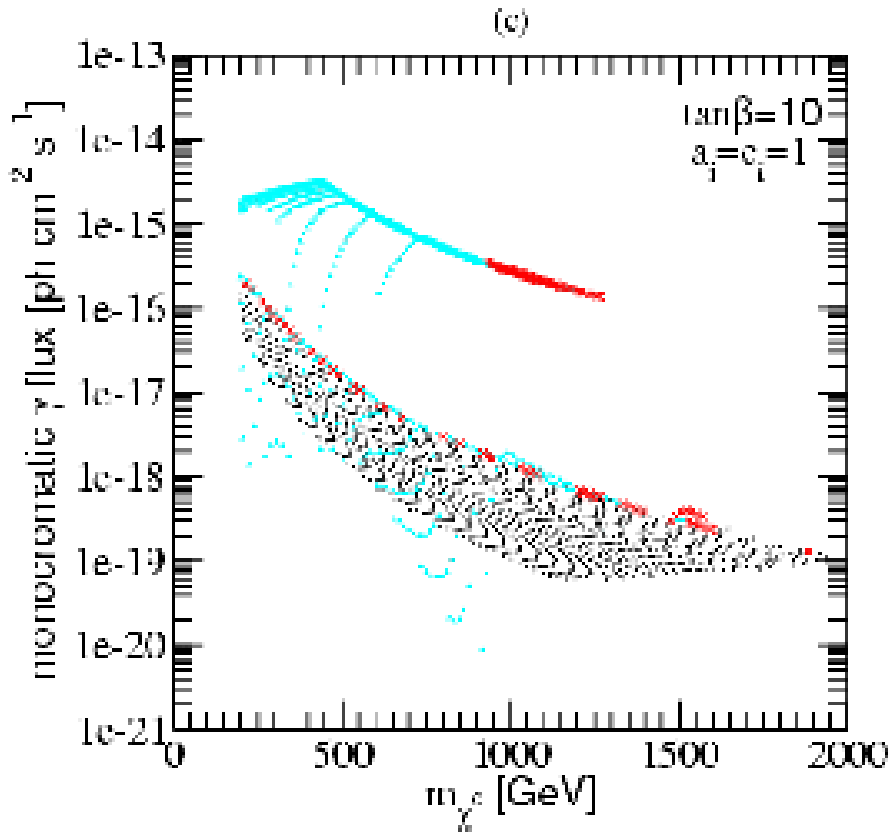}\hskip 0.5cm
\includegraphics[height=6cm,width=6cm]{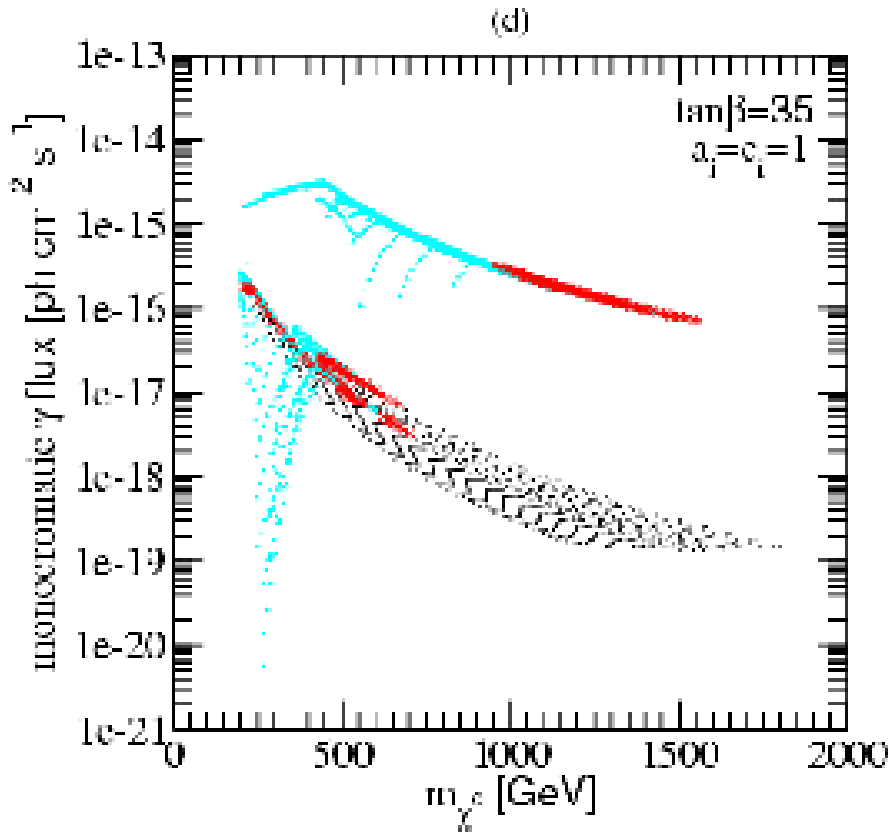}
\end{center}
\vskip -0.5cm \caption{Continuum (a and b) and monochromatic (c
and d) gamma ray flux from the Galactic Center vs. $m_\chi$ for
the case with $a_i=c_i=1$ and $\rm{tan}\beta=10$ or $35$.}
\label{fig:fluxgam-tb35}
\end{figure}
%

Gamma rays induced by  neutralino annihilation in  Galactic Center
might provide an indirect detection  of neutralino DM.
Fig.~\ref{fig:fluxgam-tb35} shows the predicted continuum (a and
b) and monochromatic (c and d) gamma ray fluxes from the Galactic
Center  as a function of the LSP neutralino mass for the case with
$a_i=c_i=1$. Here we chose the same halo density profile as the
previous section, giving $\bar J(\Delta\Omega=10^{-3} {\rm sr})
\sim 30$. The red points in the figures give the WMAP DM density,
while the cyan and the rest give $\Omega_\chi h^2<0.085$ and
$\Omega_\chi h^2>0.119$, respectively, under the assumption of
pure thermal production. Again the WMAP points have two distinct
branches, the Bino-branch and the Higgsino-branch. 
Including the cyan points
giving smaller thermal relic DM density, the case with $a_i=c_i$
can give a continuum
gamma ray flux up to $2\times 10^{-11}  {\rm cm^{-2} s^{-1}}$ and
$10^{-10} {\rm cm^{-2} s^{-1}}$ for $\tan\beta = 10$ and 35,
respectively.
 This maximum flux of the continuum gamma rays barely touch
the expected reach of GLAST. However, the real gamma ray flux can
be much bigger than these predictions if the actual halo density
profile is denser  than the assumed profile. For Higgsino LSP,
unsuppressed annihilation into W or Z boson pair is the major
source of continuum gamma rays. As can be noticed from
Fig.~\ref{fig:fluxgam-tb35}, for some parameter values, the gamma
ray flux from Bino LSP is largely enhanced by the pseudoscalar
Higgs resonance effect.

\begin{figure}[ht!]
\begin{center}
\includegraphics[height=6cm,width=6cm]{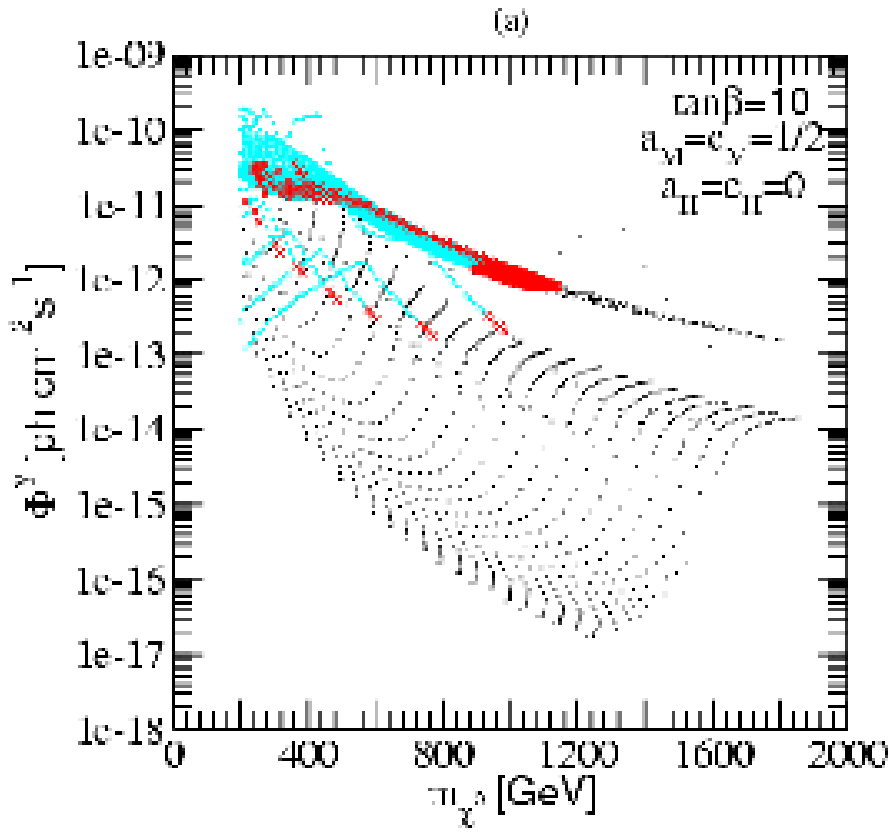}\hskip 0.5cm
\includegraphics[height=6cm,width=6cm]{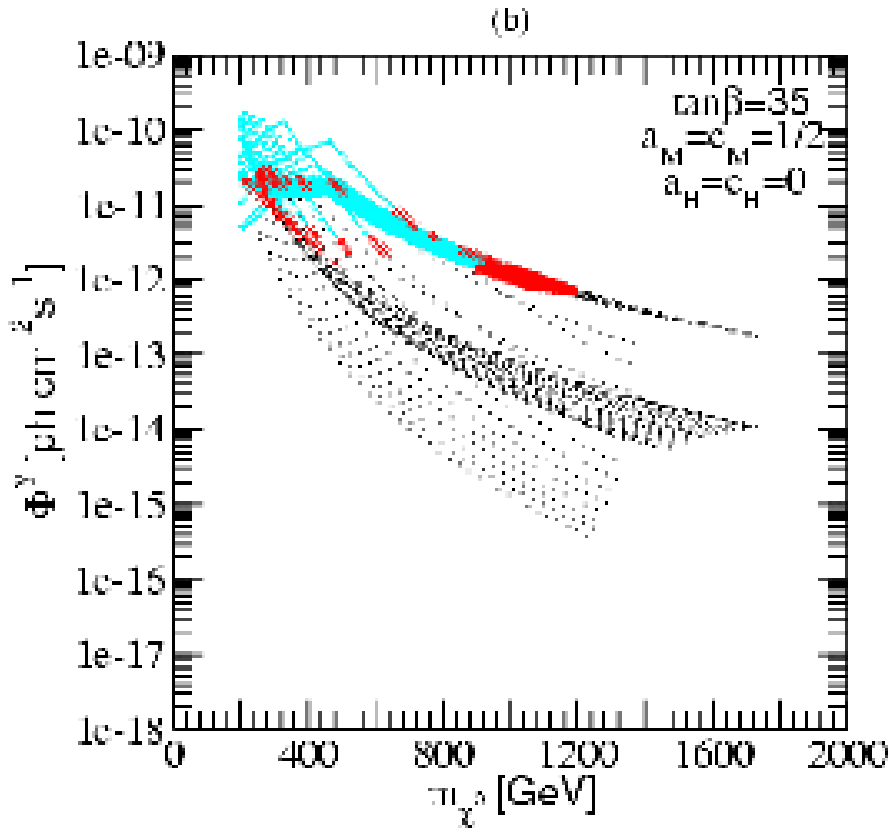}\vskip 1.0cm
\includegraphics[height=6cm,width=6cm]{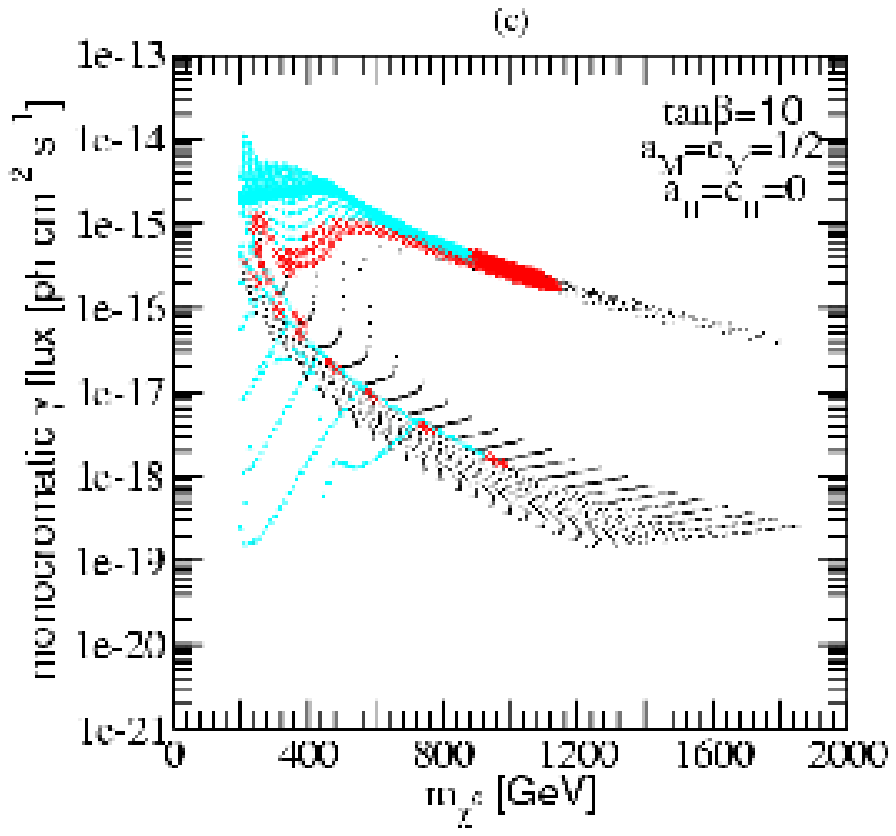}\hskip 0.5cm
\includegraphics[height=6cm,width=6cm]{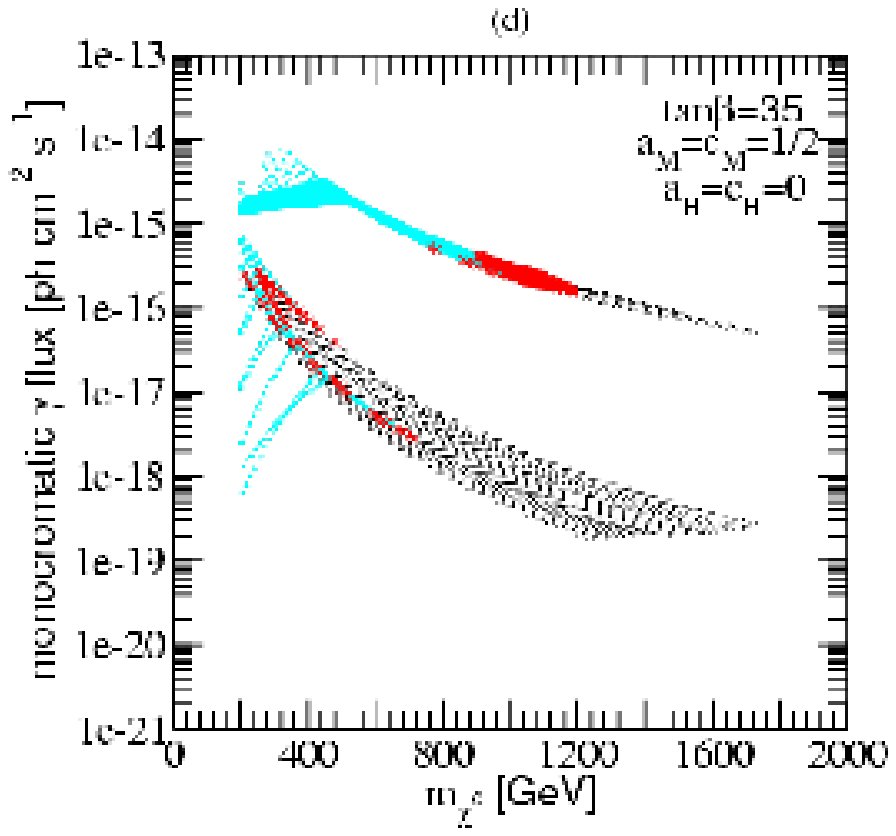}
\end{center}
\vskip -0.4cm \caption{Continuum (a and b) and monochromatic (c
and d) gamma ray flux from the Galactic Center vs. $m_\chi$ for
the case in which $a_{\rm H}=c_{\rm H}=0$, $a_{\rm M}=c_{\rm
M}=1/2$, and $\rm{tan}\beta=10$ or 35.} \label{fig:fluxgam-nzmw3}
\end{figure}


In fact, it is quite nontrivial to discriminate the continuum
gamma rays produced by neutralino annihilation from the diffuse
galactic gamma ray backgrounds. On the other hand, the
monochromatic gamma ray line from $\chi\chi \rightarrow
\gamma\gamma$ or $\gamma Z$ can be considered as a 'smoking gun'
signal of WIMP dark matter. Figs.~\ref{fig:fluxgam-tb35}.c and
\ref{fig:fluxgam-tb35}.d  show the gamma ray line flux produced by
neutralino pair annihilation  in Galactic Center for the case with
$a_i=c_i$. One can notice that there is a clear distinction
between the Bino and Higgsino LSP regions. The gamma ray line flux
ranges from $10^{-19}$ to $10^{-16}\, {\rm cm^{-2} s^{-1}}$ for
the Bino LSP branch of WMAP points,  while it ranges from
$10^{-16}$ to $10^{-15}\, {\rm cm^{-2} s^{-1}}$ for the Higgsino
LSP branch. For Higgsino-like LSP,  the gamma ray line flux comes
dominantly from the $W^\pm \chi_1^\mp$ loop diagrams resulting in
a large cross section for $\chi\chi \rightarrow \gamma\gamma$ or
$\gamma Z$ \cite{higgsinoDM}. While GLAST will probe the photon
energies only up to 300 GeV with a low energy threshold,
Atmospheric Cherenkov Telescopes (ACT) such as H.E.S.S. will be
able to cover higher photon energy ranges and probe the gamma ray
flux down to $10^{-14} {\rm cm^{-2} s^{-1}}$. The predicted
monochromatic fluxes in Figs.~\ref{fig:fluxgam-tb35}.c and
\ref{fig:fluxgam-tb35}.d  are still below this sensitivity limit.
However, as we have stressed,  these results are based on a rather
conservative halo density profile. In view of that the predicted
flux can increase even by a factor of $10^4$ if one uses an
extreme halo model like the spiked profile, the monochromatic
gamma ray signal for the Higgsino dark matter might be measurable
in case of a cuspy halo density profile.

As we have anticipated, other values of $(a_i,c_i)$ specified in
(\ref{4choices}) allow a mixed Bino-Higgsino LSP over a
significant fraction of parameter space. It is thus expected that
those other cases can give a larger gamma ray flux compared to the
case with $a_i=c_i=1$. In Fig.~\ref{fig:fluxgam-nzmw3}, we
depicted the results for the case with $a_{\rm H}=c_{\rm H}=0$ and
$a_{\rm M}=c_{\rm M}=1/2$. Indeed, this case gives a  larger flux,
although not dramatically different. The red WMAP points can give
a continuum gamma ray flux up to $3 \times 10^{-11} {\rm cm^{-2}
s^{-1}}$, while the cyan points giving  smaller thermal relic DM
density can reach up to $2\times 10^{-10} {\rm cm^{-2} s^{-1}}$.
The maximal flux of monochromatic gamma ray is about $10^{-15}
{\rm cm^{-2} s^{-1}}$ for the red WMAP points and about  $7 \times
10^{-15} {\rm cm^{-2} s^{-1}}$ for the cyan points. Again, these
results are obtained for the conservative halo density profile
giving $\bar J(\Delta\Omega=10^{-3} {\rm sr}) \sim 30$. The real
gamma ray flux can be significantly bigger than these predictions
if the actual halo density profile is denser  than the assumed
profile.

\section{Conclusions}

In this paper, we have examined the prospect of  neutralino dark
matter in mirage mediation scenario of SUSY breaking in which soft
masses receive comparable contributions from modulus mediation and
anomaly mediation. Depending upon the model parameters, especially
the anomaly to modulus mediation ratio, the nature of the lightest
neutralino changes from Bino-like to Higgsino-like via
Bino-Higgsino mixing region. For Bino-like LSP, the conventional
thermal production mechanism can give a right amount of relic DM
density, i.e. the WMAP observation $0.085 < \Omega_{DM}h^2 <
0.119$, through the stop/stau-neutralino coannihilation process or
the pseudo-scalar Higgs resonance effect. In overall, compared to
the mSUGRA scenario, a significantly larger fraction of the
parameter space  can give the WMAP DM density under the assumption
of thermal production, while satisfying all known phenomenological
constraints. This is partly because the lightest neutralino is a
mixed Bino-Higgsino over a sizable fraction of the parameter
space.

We also studied the detection possibilities of neutralino dark
matter in mirage mediation.  For the parameter region giving the
WMAP density of Bino-like or Higgsino-like LSP, direct detection
via elastic scattering between neutralino DM and nuclear target
turns out to be mostly under the sensitivity of near future
experiments. However the other parameter region giving the WMAP
density of mixed Bino-Higgsino LSP predicts typically a cross
section above the expected sensitivity limit of SuperCDMS. The
continuum and monochromatic gamma ray fluxes from neutralino
annihilation in Galactic Center have been analyzed also.
Generically, Higgsino-like LSP gives a larger gamma ray flux than
Bino-like LSP, however the continuum gamma ray flux from Bino LSP
can be significantly enhanced for some particular parameter values
due to the pseudo-scalar Higgs resonance effect. Although the
gamma ray fluxes predicted within a conservative halo model are
below the sensitivity of ongoing and planned experiments, it might
be detectable if the actual halo density  is denser  than the
conservative profile used in our analysis.

\bigskip

\acknowledgments We thank Kwang-Sik Jeong for helpful discussions
and also for clarifying various conventions for soft terms. This
work is supported by the KRF Grant KRF-2005-201-C00006 funded by
the Korean Government (K.C. and Y.S.), the KOSEF Grant
R01-2005-000-10404-0 (K.C. and Y.S.), the Center for High Energy
Physics of Kyungpook National University (K.C.), the BK21 program
of Ministry of Education (K.Y.L.), and the Astrophysical Research
Center for the Structure and Evolution of the Cosmos funded by the
KOSEF (Y.G.K.). K.O. has been supported by the grant-in-aid for
scientific research on priority areas (No. 441): "Progress in
elementary particle physics of the 21 century through discoveries
of Higgs boson and supersymmetry" (No. 16081209) from the Ministry
of Education, Culture, Sports, Science and Technology of Japan.
K.O. and Y.S. thank Yukawa Institute in Kyoto University for the
use of Altix3700 BX2 by which much of the numerical calculation
has been made. Y.S. also thanks the Particle Theory and Cosmology
Group at Tohoku University for the use of the computer facility.


\section*{Appendix A.}

\vskip 0.5cm In this appendix, we summarize  the notations and
conventions used in this paper. The quantum effective action in
$N=1$ superspace is given by \bea && \int d^4\theta
\left[-3CC^*e^{-K/3} +\frac{1}{16}\left(
G_aW^{a\alpha}\frac{D^2}{\partial^2}W^a_\alpha+{\rm
h.c.}\right)\right] +\left(\,\int d^2\theta\, C^3W+{\rm h.c.}\, \right) \nonumber \\
&=& \int d^4\theta
\left[\,-3CC^*e^{-K_0/3}+CC^*e^{-K_0/3}Z_i\Phi^*_ie^{2V_aT_a}\Phi_i
 +\frac{1}{16}\left(\,
 G_aW^{a\alpha}\frac{D^2}{\partial^2}W^a_\alpha+{\rm
 h.c.}\,\right)\,\right]\nonumber \\
 &&+\,\left(\,\int d^2\theta\,
 C^3\left[\,W_0+\frac{1}{6}\lambda_{ijk}\Phi_i\Phi_j\Phi_k\,\right]+{\rm h.c.}
\,\right)+..., \eea where the gauge kinetic terms are written as a
$D$-term operator to accommodate the radiative corrections to
gauge couplings, and the ellipsis stands for the irrelevant higher
dimensional operators. The K\"ahler potential $K$ is expanded as
\bea K=K_0(T_A,T_A^*)+Z_i(T_A,T_A^*)\Phi^*_ie^{2V_aT_a}\Phi_i+...,
\eea where $V_a$ and $\Phi_i$ denote the visible gauge and matter
superfields given by \bea
\Phi^i&=&\phi^i+\sqrt{2}\,\theta\psi^i+\theta^2F^i,\nonumber \\
V^a &=& -\theta\sigma^\mu\bar\theta A^a_\mu
-i\bar\theta^2\theta\lambda^a + i\theta^2\bar\theta\bar\lambda^a +
\frac{1}{2}\theta^2\bar\theta^2 D^a, \eea  and $T_A=(C,T)$ are the
SUSY breaking messengers including the conformal compensator
superfield $C=C_0+\theta^2F^C$ and the modulus superfield
$T=T_0+\sqrt{2}\theta\tilde{T}+\theta^2F^T$. The radiative
corrections due to renormalizable gauge and Yukawa interactions
can be encoded in the matter K\"ahler metric $Z_i$ and the gauge
coupling superfield $G_a$ which is given by
  \bea G_a\,=\,{\rm Re}(f_a)+\Delta G_a, \eea where
$f_a$ is the holomorphic gauge kinetic function and $\Delta G_a$
includes  the $T_A$-dependent radiative correction to gauge
coupling. The superpotential is expanded as \bea
W=W_0(T)+\frac{1}{6}\lambda_{ijk}(T)\Phi_i\Phi_j\Phi_k+..., \eea
where $W_0(T)$ is the modulus superpotential stabilizing $T$. Here
we do not specify the mechanism to generate the MSSM Higgs
parameters $\mu$ and $B$, and treat them as free parameters
constrained only by the electroweak symmetry breaking condition.
For a discussion of $\mu$ and $B$ in mirage mediation, see
Ref.~\cite{Choi:2005uz}.

For the canonically normalized component fields, the above
superspace action gives the following form of the running gauge
and Yukawa couplings, the supersymmetric gaugino-matter fermion
coupling ${\cal L}_{\lambda\psi}$, and the soft SUSY breaking
terms: \bea \frac{1}{g_a^2}&=& {\rm Re}(G_a),\quad y_{ijk}\, =\,
\frac{\lambda_{ijk}}{\sqrt{e^{-K_0}Z_iZ_jZ_k}},\nonumber \\
 {\cal
L}_{\lambda\psi}&=&i\sqrt{2}\left(\phi_i^* T^a\psi_i\lambda^a
-\bar\lambda^aT^a\phi_i\bar\psi_i \right), \nonumber
\\
{\cal L}_{\rm soft}&=&-m^2_i\phi^i\phi^{i*}
-\left(\,\frac{1}{2}M_a\lambda^a\lambda^a
+\frac{1}{6}A_{ijk}y_{ijk}\phi^i\phi^j\phi^k +{\rm h.c.}\right),
\eea where \bea M_a &=& F^A\partial_A\ln ({\rm Re}(G_a)),
\nonumber \\
A_{ijk} &=& -F^A\partial_A \ln\left(
\frac{\lambda_{ijk}}{e^{-K_0}Z_iZ_jZ_k}\right),
\nonumber \\
m^2_i &=&  -F^AF^{B*}\partial_A\partial_{\bar B} \ln\left(
e^{-K_0/3}Z_i\right) \eea for \bea F^T&=&
-e^{K_0/2}(\partial_T\partial_{T^*})^{-1}(D_TW_0)^*,\nonumber \\
F^C&=&m_{3/2}^*+\frac{1}{3}\partial_TK_0F^T \quad(m_{3/2}=
e^{K_0/2}W_0). \eea In the approximation ignoring the off-diagonal
components of $w_{ij}=\sum_{pq}y_{ipq}y^*_{jpq}$, the 1-loop RG
evolution of soft parameters is determined by \bea
{16\pi^2}\frac{dM_a}{d\ln\mu}&=& 2 \left[-3\,{\rm
tr}\Big(T_a^{2}({\rm Adj})\Big) +\sum_i {\rm
tr}\Big(T_a^{2}(\phi^i)\Big) \right] g^2_aM_a,
\nonumber \\
{16\pi^2}\frac{dA_{ijk}}{d\ln\mu} &=& \left[
\sum_{p,q}|y_{ipq}|^2A_{ipq} - 4 \sum_a g^2_aC_2^a(\phi^i) M_a
\right] + \Big[i \leftrightarrow j\Big] + \Big[i \leftrightarrow
k\Big],
\nonumber \\
{16\pi^2}\frac{d m^2_i}{d\ln\mu} &=&
\sum_{j,k}|y_{ijk}|^2\left(m^2_i+m^2_j+m^2_k+|A_{ijk}|^2\right)
\nonumber \\ &-& 8 \sum_a g^2_aC_2^a(\phi^i)|M_a|^2 +2g_1^2q_i
\sum_j q_j m^2_j,
\eea where the quadratic Casimir $C^a_2(\phi_i)=(N^2-1)/2N$ for a
fundamental representation $\phi_i$ of the gauge group $SU(N)$,
$C_2^a(\phi_i)=q_i^2$ for the $U(1)$ charge $q_i$ of $\phi_i$.

In mirage mediation, soft terms at $M_{GUT}$ are determined by the
modulus mediation of ${\cal O}(F^T/T)$ and the anomaly mediation
of ${\cal O}(F^C/8\pi^2 C_0)$ which are comparable to each other.
In the presence of the axionic shift symmetry \bea U(1)_T: \quad
{\rm Im}(T)+ \mbox{real constant}\eea which is broken by the
non-perturbative term in the modulus superpotential \bea
W_0=w-Ae^{-aT},\eea one can always make that $m_{3/2}$ and $F^T$
are simultaneously real. Also since $F^T/T\sim m_{3/2}/4\pi^2$, we
have \bea \frac{F^C}{C_0}= m_{3/2}\left(\,1+{\cal
O}\left(\frac{1}{8\pi^2}\right)\,\right). \eea Then, upon ignoring
the parts of ${\cal O}(F^T/8\pi^2 T)$, the resulting soft
parameters at $M_{GUT}$ are given by
\begin{eqnarray}
 M_a&=& M_0 +\frac{m_{3/2}}{16\pi^2}\,b_ag_a^2,
\nonumber \\
A_{ijk}&=&\tilde{A}_{ijk}-
\frac{m_{3/2}}{16\pi^2}\,(\gamma_i+\gamma_j+\gamma_k),
\nonumber\\
m_i^2&=& \tilde{m}_i^2-\frac{m_{3/2}}{16\pi^2}M_0\,\theta_i
-\left(\frac{m_{3/2}}{16\pi^2}\right)^2\dot{\gamma}_i,
\end{eqnarray} where \bea M_0&=&F^T\partial_T\ln{\rm Re}(f_a),
\nonumber
\\
\tilde{A}_{ijk}&\equiv& (a_i+a_j+a_k)M_0
\,=\,F^T\partial_T\ln(e^{-K_0}Z_iZ_jZ_k),\nonumber \\
\tilde{m}_i^2&\equiv& c_iM_0^2\,=\,
-|F^T|^2\partial_T\partial_{\bar{T}} \ln(e^{-K_0/3}Z_i),\eea and
  \bea b_a&=&-3{\rm tr}\left(T_a^2({\rm
Adj})\right)+\sum_i {\rm tr}\left(T^2_a(\phi_i)\right),
\nonumber \\
\gamma_i&=&2\sum_a
g_a^2C^a_2(\phi_i)-\frac{1}{2}\sum_{jk}|y_{ijk}|^2, \nonumber
\\
\theta_i&=& 4\sum_a g_a^2 C^a_2(\phi_i)-\sum_{jk}|y_{ijk}|^2
(a_i+a_j+a_k), \nonumber
\\
\dot{\gamma}_i&=&8\pi^2\frac{d\gamma_i}{d\ln\mu},\eea where
$\omega_{ij}=\sum_{kl}y_{ikl}y^*_{jkl}$ is assumed to be diagonal.
Here we have used that $\lambda_{ijk}$ are $T$-independent
constant as ensured by the axionic shift symmetry $U(1)_T$.

Let us now summarize our conventions for the MSSM. The
superpotential of canonically normalized matter superfields is
given by \bea W &=& y_DH_d\cdot QD^c+y_LH_d\cdot LE^c-y_UH_u\cdot
QU^c - \mu H_d\cdot H_u, \eea where the $SU(2)_L$ product is
$H\cdot Q=\epsilon_{ab}H^aQ^b$ with
$\epsilon_{12}=-\epsilon_{21}=1$, and color indices are
suppressed.  Then the chargino and neutralino mass matrices are
given by \bea -\frac{1}{2}\,\tilde\psi^{-T}{\cal M}_C\tilde\psi^+
-\frac{1}{2}\,\tilde\psi^{0T}{\cal M}_N\tilde\psi^0 + {\rm h.c.},
\eea where \bea {\cal M}_C &=& \left(
\begin{array}{cc}
- M_2\,\,
& g_2 \langle H^0_u \rangle \\
g_2 \langle H^0_d \rangle & \mu
\end{array}
\right),\nonumber \\
{\cal M}_N &=& \left(
\begin{array}{cccc}
-M_1 & 0 & -\frac{1}{\sqrt 2}\,g_Y \langle H^0_d \rangle
& \frac{1}{\sqrt 2}\,g_Y \langle H^0_u \rangle \\
0 & -M_2 & \frac{1}{\sqrt 2}\,g_2 \langle H^0_d \rangle
& -\frac{1}{\sqrt 2}\,g_2 \langle H^0_u \rangle \\
  -\frac{1}{\sqrt 2}\,g_Y \langle H^0_d \rangle
& \frac{1}{\sqrt 2}\,g_2 \langle H^0_d \rangle
& 0 & -\mu  \\
  \frac{1}{\sqrt 2}\,g_Y \langle H^0_u \rangle
& -\frac{1}{\sqrt 2}\,g_2 \langle H^0_u \rangle & -\mu & 0
\end{array}
\right), \eea in the field basis \bea \tilde\psi^{+T} &=&
-i\left(\tilde W^+,\, i\tilde H^+_u \right), \quad \tilde\psi^{-T}
\,=\, -i\left(\tilde W^-,\, i\tilde H^-_d \right),
\nonumber \\
\tilde\psi^{0T} &=& -i\left( \tilde B,\,\tilde W^3,\, i\tilde
H^0_d,\,i\tilde H^0_u \right), \eea for $\tilde W^{\pm}=(\tilde
W^1\mp i \tilde W^2)/\sqrt 2.$

The one-loop beta function coefficients $b_a$ and anomalous
dimension $\gamma_i$ in the MSSM are given by \bea b_3&=&-3, \qquad
b_2=1,\qquad b_1=\frac{33}{5},
\nonumber \\
\gamma_{H_u} &=& \frac{3}{2}g_2^2+\frac{1}{2}g_Y^2 -3y_t^2,
\nonumber \\
\gamma_{H_d} &=& \frac{3}{2}g_2^2+\frac{1}{2}g_Y^2 - 3 y_b^2 -  y_\tau^2
\nonumber \\
\gamma_{Q_a} &=& \frac{8}{3} g_3^2 + \frac{3}{2} g_2^2
                +\frac{1}{18} g_Y^2 - (y_t^2 + y_b^2) \delta_{3a},
\nonumber \\
\gamma_{U_a} &=& \frac{8}{3} g_3^2  + \frac{8}{9} g_Y^2
                - 2 y_t^2 \delta_{3a},
\nonumber \\
\gamma_{D_a} &=& \frac{8}{3} g_3^2 + \frac{2}{9} g_Y^2
                - 2 y_b^2 \delta_{3a},
\nonumber \\
\gamma_{L_a} &=& \frac{3}{2} g_2^2 + \frac{1}{2} g_Y^2
                - y_\tau^2 \delta_{3a},
\nonumber \\
\gamma_{E_a} &=& 2 g_Y^2 - 2 y_\tau^2 \delta_{3a}, \eea where
$g_2$ and $g_Y=\sqrt{3/5}g_1$ denote the $SU(2)_L$ and $U(1)_Y$
gauge couplings.
 The
$\theta_i$ and $\dot{\gamma}_i$ which determine the soft scalar
masses at $M_{GUT}$ are given by \bea \theta_{H_u} &=&
3g_2^2+g_Y^2 -6y_t^2(a_{H_u}+a_{Q_3}+a_{U_3}), \nonumber \\
\theta_{H_d} &=& 3g_2^2+g_Y^2 - 6y_b^2(a_{H_d}+a_{Q_3}+a_{D_3}) -
2y_\tau^2(a_{H_d}+a_{L_3}+a_{E_3})
\nonumber \\
\theta_{Q_a} &=& \frac{16}{3} g_3^2 + 3 g_2^2
                +\frac{1}{9} g_Y^2 - 2\Big(y_t^2(a_{H_u}+a_{Q_3}+a_{U_3}) + y_b^2(a_{H_d}+a_{Q_3}+a_{D_3})\Big) \delta_{3a},
\nonumber \\
\theta_{U_a} &=& \frac{16}{3} g_3^2  + \frac{16}{9} g_Y^2
                - 4y_t^2(a_{H_u}+a_{Q_3}+a_{U_3}) \delta_{3a},
\nonumber \\
\theta_{D_a} &=& \frac{16}{3} g_3^2 + \frac{4}{9} g_Y^2
                - 4y_b^2(a_{H_d}+a_{Q_3}+a_{D_3}) \delta_{3a},
\nonumber \\
\theta_{L_a} &=& 3 g_2^2 + g_Y^2
                - 2y_\tau^2 (a_{H_d}+a_{L_3}+a_{E_3})\delta_{3a},
\nonumber \\
\theta_{E_a} &=& 4 g_Y^2 - 4 y_\tau^2(a_{H_d}+a_{L_3}+a_{E_3})
\delta_{3a}, \eea and \bea \dot\gamma_{H_u} &=& \frac{3}{2} g_2^4
+ \frac{11}{2} g_Y^4
                     - 3 y_t^2 b_{y_t},
\nonumber \\
\dot\gamma_{H_d} &=& \frac{3}{2} g_2^4 + \frac{11}{2} g_Y^4
                    - 3 y_b^2 b_{y_b} - y_\tau^2 b_{y_\tau},
\nonumber \\
\dot \gamma_{Q_a} &=&  -8 g_3^4 + \frac{3}{2} g_2^4 + \frac{11}{18} g_Y^4
                  -(y_t^2 b_{y_t} + y_b^2  b_{y_b}) \delta_{3a},
\nonumber \\
\dot\gamma_{U_a} &=& - 8 g_3^4  +  \frac{88}{9} g_Y^4
                   - 2 y_t^2 b_{y_t} \delta_{3a},
\nonumber \\
\dot\gamma_{D_a} &=& - 8 g_3^4 + \frac{22}{9} g_Y^4
                     - 2 y_b^2 b_{y_b} \delta_{3a},
\nonumber \\
\dot\gamma_{L_a} &=& \frac{3}{2}g_2^4 + \frac{11}{2} g_Y^4
                     - y_\tau^2 b_{y_\tau} \delta_{3a},
\nonumber \\
\dot\gamma_{E_a} &=& 22 g_Y^4 - 2 y_\tau^2 b_{y_\tau} \delta_{3a},
\eea
where
\bea
b_{y_t} &=& - \frac{16}{3} g_3^2 - 3 g_2^2 - \frac{13}{9} g_Y^2
             + 6 y_t^2 + y_b^2,
\nonumber \\
b_{y_b} &=& - \frac{16}{3} g_3^2 - 3 g_2^2 - \frac{7}{9} g_Y^2
             + y_t^2 + 6 y_b^2 + y_\tau^2,
\nonumber \\
b_{y_\tau} &=& - 3 g_2^2 - 3 g_Y^2 + 3 y_b^2  + 4 y_\tau^2. \eea


\begin{thebibliography}{999}

\bibitem{Nilles:1983ge}
  H.~P.~Nilles,
  Phys.\ Rept.\  {\bf 110}, 1 (1984);
%
  H.~E.~Haber and G.~L.~Kane,
  Phys.\ Rept.\  {\bf 117}, 75 (1985).


\bibitem{lspdm}
For a review, see G.~Jungman, M.~Kamionkowski and K.~Griest,
Phys. Rep. {\bf 267}, 195 (1996);
G.~Bertone, D.~Hooper and J.~Silk, Phys. Rep. {\bf 405}, 279 (2005)
[arXiv:hep-ph/0404175]

\bibitem{wmap}
D.~N.~Spergel {\it et al.} [arXiv:astro-ph/0603449].



\bibitem{gkp}
S. B. Giddings, S. Kachru and  J. Polchinski, Phys. Rev. {\bf
D66}, 106006 (2002) [arXiv: hep-th/0105097].


\bibitem{kklt}
S.~Kachru, R.~Kallosh, A.~Linde and S.~P.~Trivedi,
Phys.\ Rev.\ D {\bf 68}, 046005 (2003) [arXiv:hep-th/0301240].

\bibitem{choi1}
  K.~Choi, A.~Falkowski, H.~P.~Nilles, M.~Olechowski and S.~Pokorski,
  JHEP {\bf 0411}, 076 (2004)
  [arXiv:hep-th/0411066];
  K.~Choi, A.~Falkowski, H.~P.~Nilles and M.~Olechowski,
Nucl. Phys. {\bf B718}, 113 (2005) [arXiv:hep-th/0503216].



\bibitem{modulus}
V.~S.~Kaplunovsky and J.~Louis,
Phys.\ Lett.\ B {\bf 306}, 269 (1993) [arXiv: hep-th/9303040];
A.~Brignole, L.~E.~Ibanez and C.~Munoz,
Nucl.\ Phys.\ B {\bf 422}, 125 (1994) [Erratum-ibid.\ B {\bf 436},
747 (1995)] [arXiv: hep-ph/9308271].



\bibitem{anomaly}
  L.~Randall and R.~Sundrum,
  Nucl.\ Phys.\ B {\bf 557}, 79 (1999)
  [arXiv: hep-th/9810155];
  G.~F.~Giudice, M.~A.~Luty, H.~Murayama and R.~Rattazzi,
  JHEP {\bf 9812}, 027 (1998)
  [arXiv: hep-ph/9810442];
J. A. Bagger, T. Moroi and E. Poppitz, JHEP {\bf 0004}, 009 (2000)
[arXiv: hep-th/9911029];
  P.~Binetruy, M.~K.~Gaillard and B.~D.~Nelson,
  Nucl.\ Phys.\ B {\bf 604}, 32 (2001)
  [arXiv: hep-ph/0011081].

\bibitem{Choi:2005uz}
  K.~Choi, K.~S.~Jeong and K.~i.~Okumura,
  JHEP {\bf 0509}, 039 (2005)
  [arXiv:hep-ph/0504037].


\bibitem{ratz} O. L.-Brito, J. Martin, H. P. Nilles and M. Ratz,
hep-th/0509158



\bibitem{endo05}
M.~Endo, M.~Yamaguchi and K.~Yoshioka, Phys. Rev. {\bf D 72},
015004 (2005)[arXiv:hep-ph/0504036].



\bibitem{falkowski05}
A.~Falkowski, O.~Lebedev and Y.~Mambrini,
JHEP {\bf 0511}, 034 (2005)[arXiv: hep-ph/0507110].




\bibitem{baer06}
H.~Baer, E.-K.~Park, X.~Tata and T.~T. Wang
hep-ph/0604253.

\bibitem{baer}
 H. Baer, E.-K. Park, X. Tata and T.~T. Wang,
hep-ph/0607085.

\bibitem{yama}
M.~Endo, K.~Hamaguch and F.~Takahashi, Phys. Rev. Lett. {\bf
96} (2006) 211301 [arXiv:hep-ph/0602061]; S.~Nakamura and
M.~Yamaguchi, Phys. Lett. B {\bf 638}, 389 (2006)
[arXiv:hep-ph/0602081]; T. Asaka, S. Nakamura and M. Yamaguchi,
Phys. Rev. {\bf D74}, 023520 (2006) [arXiv:hep-ph/0604132]; M.
Dine, R. Kitano, A. Morisse and Y. Shirman, Phys. Rev. {\bf D73},
123518 (2006) [arXiv:hep-ph/0604140]

\bibitem{ellis06}
J.~Ellis, K.~Olive and P.~Sandick,
hep-ph/0607002.

\bibitem{stewart} D. H. Lyth and E. D. Stewart,
Phys.Rev. {\bf D 53}, 1784 (1996) [arXiv:hep-ph/9510204]; Phys.
Rev. Lett. {\bf 75}, 201 (1995)  [arXiv:hep-ph/9502417]

\bibitem{nonthermal}
T.~Moroi and L.~Randall, Nucl. Phys. B {\bf 570}, 455 (2000) [arXiv:hep-ph/9906527];

\bibitem{hdkim}
R. Dermisek and H. D. Kim, Phys. Rev. Lett. {\bf 96}, 211803
(2006) [arXiv:hep-ph/0601036];
R. Dermisek, H. D. Kim and I.-W.  Kim, hep-ph/0607169.

\bibitem{munoz}
J. A. Casas, A. Lleyda and C. Munoz, Nucl. Phys. {\bf B471}, 3
(1996) [arXiv:hep-ph/9507294]

\bibitem{kuzenko}
A. Kusenko, P. Langacker and G. Segre, Phys. Rev. {\bf D54}, 5824
(1996) [arXiv:hep-ph/9602414]; A. Kusenko and P. Langacker, Phys.
Lett. B {\bf 391}, 29 (1997) [arXiv:hep-ph/9608340].

\bibitem{riotto}
A. Riotto and E. Roulet, Phys. Lett. B {\bf 377}, 60 (1996)
[arXiv:hep-ph/9512401]



\bibitem{choi3}
  K.~Choi and K.~S.~Jeong,
  arXiv:hep-th/0605108.
\bibitem{hebecker}
 F. Brummer, A. Hebecker and M. Trapletti, hep-th/0605232.




\bibitem{susycp}
K.~Choi,
Phys.\ Rev.\ Lett.\  {\bf 72}, 1592 (1994)
[arXiv:hep-ph/9311352].

\bibitem{luty}
 M. Luty and R. Sundrum, Phys. Rev. {\bf D62}, 035008 (2000) [arXiv:hep-th/9910202];
 Phys. Rev. {\bf D64}, 065012 (2001)
[arXiv:hep-th/0012158].



\bibitem{ck}
  K.~Choi and J.~E.~Kim,
  Phys.\ Lett.\ B {\bf 165}, 71 (1985);
%
  L.~E.~Ibanez and H.~P.~Nilles,
  Phys.\ Lett.\ B {\bf 169}, 354 (1986); T.~Banks and M.~Dine,
  Nucl.\ Phys.\ B {\bf 479}, 173 (1996)
  [arXiv:hep-th/9605136];
  K.~Choi,
  Phys.\ Rev.\ D {\bf 56}, 6588 (1997)
  [arXiv:hep-th/9706171].
%
%


\bibitem{lust} D. Lust, P. Mayr, R. Richter and S. Stieberger,
Nucl. Phys. {\bf B696}, 205 (2004) [arXiv:hep-th/0404134];
M. Bertolini et al., Nucl. Phys. {\bf B743}, 1 (2006) [arXiv:hep-th/0512067].


\bibitem{abe}
H.~Abe, T.~Higaki and T.~Kobayashi,
D {\bf 73}, 046005 (2006)
[arXiv:hep-th/0511160].


\bibitem{tevmirage}
K. Choi, K. S. Jeong, T. Kobayashi and Ken-ichi Okumura, Phys.
Lett. B {\bf 633}, 355 (2006) [arXiv:hep-ph/0508029]; R. Kitano
and Y. Nomura, Phys. Lett. B {\bf 631}, 58 (2005)
[arXiv:hep-ph/0509039]; O.~Lebedev, H. P. Nilles and  M. Ratz,
arXiv:hep-ph/0511320;
  A.~Pierce and J.~Thaler,
  arXiv:hep-ph/0604192.




\bibitem{g-2}
G.~W.~Bennett {\it et al.}, [Muon g-2 Collaboration], Phys. Rev. Lett. {\bf 92},
161802 (2004) [arXiv:hep-ex/0401008];
M.~Davier, S.~Eidelman, A.~Hocker and Z.~Zhang,
Eur. Phys. J. C {\bf 31} (2003) [arXiv:hep-ph/0308213];
K.~Hagiwara, A.~D.~Martin, D.~Nomura and T.~Teubner, Phys. Rev. D {\bf 69},
093003 (2004) [arXiv:hep-ph/0312250];
J.~F.~de~Troconiz and F.~J.~Yndurain, arXiv:hep-ph/0402285;
K.~Melnikov and A.~Vainshtein, arXiv:hep-ph/0312226;
M.~Passera, arXiv:hep-ph/0411168.

\bibitem{darksusy}
P.~Gondolo, J.~Edsjo, L.~Bergstrom, P.~Ullio and E.~A.~Baltz,
arXiv:astro-ph/0012234;
P.~Gondolo, J.~Edsjo, P.~Ullio, L.~Bergstrom, M.~Schelke and E.~A.~Baltz,
JCAP 0407 (2004) 008 [arXiv:astro-ph/0406204].

\bibitem{goodman}
M.~W.~Goodman and E.~Witten, Phys. Rev. {\bf D 31} (1985) 3059.


\bibitem{supercdms}
SuperCDMS, arXiv:astro-ph/0503583.

\bibitem{moore}
J.~Diemand, B.~Moore and J.~Stadel,
     Mon. Not. Roy. Astron. Soc. {\bf 353}, 624 (2004);
B.~Moore et al., Astrophys. J. {\bf 524}, L19 (1999).

\bibitem{hess}
F.~Aharonian {\it et al.}, The HESS collaboration, astro-ph/0408145.

\bibitem{hooper}
G.~Zaharijas and D.~Hooper, Phys. Rev. {\bf D 73}, 103501 (2006).

\bibitem{coanil}
S.~Mizuta and M.~Yamaguchi,
Phys. Lett. B {\bf 298} (1993) 120 [arXiv:hep-ph/9208251];
U.~Chattopadhyay, D.~Choudhury, M.~Drees, P.~Konar and D.~Roy,
Phys. Lett. B {\bf 632} (2006) 114 [arXiv:hep-ph/0508098].


\bibitem{higgsinoDM}
L.~Bergstrom, P.~Ullio and J.H.~Buckley, Astropart Phys. {\bf 9}, 137 (1998);
P.~Ullio and L.~Bergstrom, Phys. Rev. D {\bf 57}, 1962 (1998).
















%
%


\end{thebibliography}
\end{document}